\documentclass[lettersize,journal]{IEEEtran}
\usepackage{amsmath,amsfonts}
\usepackage{algorithmic}
\usepackage[linesnumbered,lined,boxed,commentsnumbered,ruled,vlined]{algorithm2e}
\usepackage{array}
\usepackage[caption=false,font=normalsize,labelfont=sf,textfont=sf]{subfig}
\usepackage{textcomp}
\usepackage{stfloats}
\usepackage{url}
\usepackage{verbatim}
\usepackage{graphicx}
\usepackage{cite}
\usepackage{amsmath}
\usepackage{amssymb}
\begin{document}

\title{Multiperson Detection and Vital-Sign Sensing Empowered by Space-Time-Coding RISs}

\author{Xinyu Li,~\IEEEmembership{Member,~IEEE}, Jian Wei You,~\IEEEmembership{Senior Member,~IEEE}, Ze Gu, Qian Ma,~\IEEEmembership{Member,~IEEE},\\ Jingyuan Zhang, Long Chen, and Tie Jun Cui,~\IEEEmembership{Fellow,~IEEE}
	% <-this % stops a space
	\thanks{Xinyu Li, Jian Wei You, Ze Gu, Qian Ma, Jingyuan Zhang, Long Chen, and Tie Jun Cui are with the State Key Laboratory of Millimeter Wave, Southeast University, Nanjing 210096, China (E-mail: \{xinyuli, jvyou, guze, maqian, zhangjingyuan, 220220703, tjcui\}@seu.edu.cn). \textit{(Corresponding author: Jian Wei You)}.}% <-this % stops a space
%	\thanks{Shi Jin is with the National Mobile Communications Research Laboratory, Southeast University, Nanjing 210096, China (E-mail: jinshi@seu.edu.cn). 
\thanks{This work was supported in part by the National Key Research and Development Program of China under Grant 2023YFB3813100, in part by the Postdoctoral Innovation Talents Support Program under Grant BX20230066, in part by the Jiangsu Planned Projects for Postdoctoral Research Fund under Grant 2023ZB318, in part by the National Natural Science Foundation of China under Grant 62288101, 6504009769, 62101124, 62371131, and 62371132, in part by the Natural Science Foundation of Jiangsu Province under Grant BK20230820 and BK20210209, in part by Fundamental Research Funds for the Central Universities under Grant 2242023K5002, in part by the China Postdoctoral Science Foundation under Grant 2021M700761 and 2022T150112, and in part by the Start-up Research Fund of Southeast University under Grant RF1028623244.}}

% The paper headers

% Remember, if you use this you must call \IEEEpubidadjcol in the second
% column for its text to clear the IEEEpubid mark.

\maketitle

\begin{abstract}
Passive human sensing using wireless signals has attracted increasing attention due to its superiorities of non-contact and robustness in various lighting conditions. However, when multiple human individuals are present, their reflected signals could be intertwined in the time, frequency and spatial domains, making it challenging to separate them. To address this issue, this paper proposes a novel system for multiperson detection and monitoring of vital signs (i.e., respiration and heartbeat) with the assistance of space-time-coding (STC) reconfigurable intelligent metasurfaces (RISs). Specifically, the proposed system scans the area of interest (AoI) for human detection by using the harmonic beams generated by the STC RIS. Simultaneously, frequency-orthogonal beams are assigned to each detected person for accurate estimation of their respiration rate (RR) and heartbeat rate (HR). Furthermore, to efficiently extract the respiration signal and the much weaker heartbeat signal, we propose an improved variational mode decomposition (VMD) algorithm to accurately decompose the complex reflected signals into a smaller number of intrinsic mode functions (IMFs). We build a prototype to validate the proposed multiperson detection and vital-sign monitoring system. Experimental results demonstrate that the proposed system can simultaneously monitor the vital signs of up to four persons. The errors of RR and HR estimation using the improved VMD algorithm are below 1 RPM (respiration per minute) and 5 BPM (beats per minute), respectively. Further analysis reveals that the flexible beam controlling mechanism empowered by the STC RIS can reduce the noise reflected from other irrelative objects on the physical layer, and improve the signal-to-noise ratio of echoes from the human chest.
 
\end{abstract}

\begin{IEEEkeywords}
Radio-frequency-based multiperson sensing, contactless physiological sensing, space-time-coding, reconfigurable intelligent surface (RIS), variational mode decomposition (VMD).
\end{IEEEkeywords}

\section{Introduction}
Vital signs are critical indicators for assessing an individual's general health state and identifying various diseases. For instance, vital signs could be employed for sleep quality estimation and further diagnosing sleep apnea \cite{8331078,carter2023deep}. The monitoring of respiration aids in the initial diagnosis of asthma \cite{9967964}. Traditional vital-sign monitoring utilizes Photoplethysmography (PPG) and Electrocardiogram (ECG) sensors to measure human respiration rate (RR) and heartbeat rate (HR) \cite{da2018internet}. Nevertheless, these sensors need to be worn and may impose a burden on the subjects. Alternatively, wireless signals offer a contact-free and cost-effective solution for vital-sign monitoring, garnering significant interest recently \cite{10.1145/3411816,9217947,10.1145/3377165}.

vital-sign monitoring using wireless signals is realized by capturing the tiny variations in reflected signals (e.g., radio absorption, scattering, and polarization) caused by the chest movements of human individuals. Commonly employed mediums for wireless human sensing include WiFi and portable radar \cite{9941043}. For instance, Ali et al. \cite{9423591} quantified the dynamics of respiration and body motions on WiFi channel state information (CSI) data, and then proposed the respiration and body-motion tracking algorithm to monitor sleeping vital signs. Liu et al. \cite{liu2023mmrh} adopted a frequency-modulated continuous-wave (FMCW) millimeter-wave (mm-wave) radar for vital-sign monitoring. They introduced a zero-attracting sign exponentially forgetting least mean square (ZA-SEFLMS) algorithm to construct a high-resolution sparse spectrum, enabling accurate and reliable estimation of RR and HR.

While current wireless signal-based vital-sign monitoring approaches have made significant strides, they are primarily suitable for single-person sensing. The extension of these methods to multi-person sensing presents notable challenges. Specifically, when there are multiple persons in the area of interest (AoI), there would be numerous multi-path components, introducing considerable interference to the echoes from the target individual. Furthermore, the signals reflected by different persons may be entangled together in the time, frequency and spatial domains, making it challenging to separate them. To deal with these challenges, Chian et al. \cite{9217947} proposed a successive signal cancellation algorithm and employed a single-antenna, narrow-band Doppler radar to monitor the vial signs of multiple persons. Mercuri et al. \cite{mercuri2019vital} introduced the time-range-speed map to distinguish between different individuals, and extracted phase information in the corresponding range bins to separate the physiological signals of the detected persons. Meanwhile, Zheng et al. \cite{zheng2020v2ifi} leveraged the high temporal resolution of ultra-wideband impulse radio to effectively differentiate between multiple persons based on their relative distances.% However, the mixed echoes signals from diverse human targets are generally disentangled at the receiving terminal by elaborating some signal processing algorithms.

In this paper, we employ the space-time-coding (STC) reconfigurable intelligent metasurface (RIS) for multiperson detection and vital signs (i.e., respiration and heartbeat) monitoring. RISs are two-dimensional (2D) artificial surfaces with reconfigurable periodic or quasi-periodic meta-atoms in subwavelength scale \cite{ma2020information}. By tailoring the electromagnetic (EM) properties of each meta-atom, the metasurface can flexibly manipulate the phase, amplitude, and polarization of the incident EM waves and consequently shape the wavefronts in programmable manner. To dynamically control metasurfaces and achieve real-time EM manipulation, a series of active technologies, including PIN diode, semiconductor, and microfluid, have been adopted for discretizing the meta-atoms into finite types and realizing electromagnetic response regulations with digital codes.

Hitherto the vast majority of metasurfaces for RF-based human sensing are based on the space-domain-coding (SDC) pattern and utilized to manipulate the spatial propagation of sensing signals \cite{li2019intelligent,hu2021metasensing,9938373,8995564}. However, although the SDC RIS can modulate the signal beams into diverse propagation directions, the echo signals from different individuals arriving at the receiving terminal are still entangled in the time, frequency and spatial domains, making it challenging to separate these signals. Compared with the SDC RISs, the STC RISs can manipulate EM waves in both the harmonic distribution (frequency domain) and the propagation direction (spatial domain) \cite{zhang2018space} by jointly encoding the parameters (e.g., reflection amplitudes and phases) of metasurfaces in time and space. Taking advantage of this characteristics, in this paper, we employ the STC RIS for the passive multi-person sensing task. With the assistance of the STC RIS, the proposed passive human sensing system could separate the echo signals from different individuals at the physical layer, without using complicated signal processing algorithms.

The advantages of employing STC RISs for human sensing can be summarized as follows. Firstly, the RISs enable the customization of wireless propagation channels, thereby enhancing sensing capabilities and enabling non-line-of-sight (NLoS) sensing. Secondly, the RISs aid in focusing sensing signals on a specific AoI, effectively mitigating interference from other irrelevant regions and significantly improving the signal-to-noise ratio (SNR) of reflected signals. Moreover, STC RISs can break through the spatial resolution limitation of the existing wireless sensing systems by generating a series of harmonic beams with diverse frequencies and propagation directions, empowering passive multi-person sensing \cite{passive}.

However, as the chest displacement caused by heartbeat is much smaller than that caused by respiration, the weaker heartbeat signal is susceptible to be overwhelmed by the respiration signal and background clutter, making it difficult to achieve accurate and reliable HR estimation. Though some mode decomposition algorithms, such as empirical mode decomposition (EMD) \cite{motin2019selection} and variational mode decomposition (VMD) \cite{dragomiretskiy2013variational}, have been utilized to extract narrow-band vital signs signals, the efficacy of these traditional algorithms is limited for accurate HR and RR estimation. Specifically, the conventional VMD algorithm may treat heartbeat signal as higher harmonics of respiration signals, complicating the extraction of heartbeat signals. Besides, the performance of these conventional algorithms is significantly impacted by the preset number of intrinsic mode functions (IMF). In other words, when the number of IMFs is insufficient, VMD is difficult or even unable to extract the weak heartbeat signals. To deal with these issues, we propose an improved VMD algorithm for the multiperson vital-sign monitoring application.

\begin{figure*}
	\centering
	\includegraphics[width=2\columnwidth]{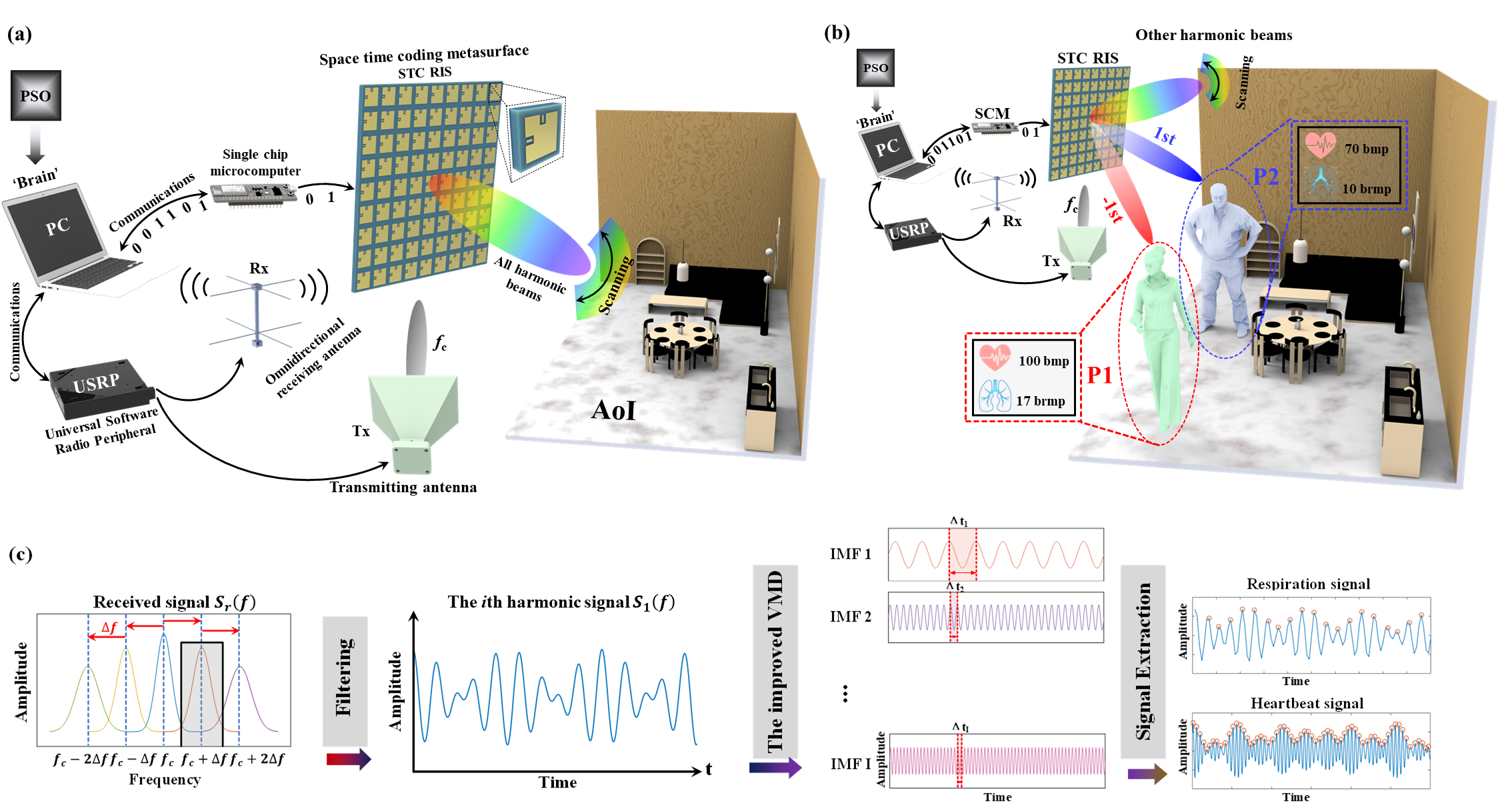}
	\caption{The workflow of the proposed multiperson detection and vital-sign monitoring system. (a) In the physical layer, the PC controls the USRP and the RIS to generate a series of harmonic beams with orthogonal frequencies for human sensing. (b) In the digital layer, the harmonic beams are scanning the AoI for target detection. Once a person is detected, a harmonic beam is assigned on this orientation for continuous vital-sign sensing, while the other beams are still utilized for human detection. (c) In the sensing layer, an improved VMD algorithm is proposed to extract respiration signals and heartbeat signals from the reflected echoes for RR and HR estimation.}
	\label{overall}
\end{figure*}

Overall, the workflow of the proposed multiperson detection and vital-sign monitoring system is shown in Fig. \ref{overall}. Specifically, an STC RIS is employed to modulate the incident wave into a set of harmonic components with diverse frequencies. Then, these frequency-orthogonal harmonic beams are controlled to scan the sensing region for human detection. Once a human individual is detected, the system would assign a harmonic beam on this orientation and monitor the vital signs of the detected person. Then, as different individuals are illuminated by the harmonic beams with different frequencies, the reflected echoes from these persons can be separated in the frequency domain. Such a sensing strategy enables to monitor multiple persons independently, significantly reducing mutual interference between the echoes from different persons. Furthermore, an improved VMD algorithm is proposed to extract physiological signals from the reflected signals and achieve accurate vital-sign sensing. Our main contributions are summarized below:

\begin{itemize}
	\item We propose a RF-based multiperson sensing system with the assistance of the STC RIS. A set of harmonic beams characterized by distinct frequencies is generated using the STC RIS, and independently controlled to illuminate multiple individuals situated at different locations for sensing purposes.  
	\item We propose a simultaneous human detection and vital-sign monitoring scheme. Specifically, the harmonic beams continuously scan the AoI and identify human presences based on two elaborated indicators. Once an individual is detected, a harmonic beam is assigned and redirected towards the individual's location for vital-sign sensing, while the remaining beams continue the scanning process.
	\item To accurately monitor human respiration and heartbeat, we propose an improved VMD algorithm, which can adaptively decompose  physiological signals from the reflected signals using a small preset number of IMFs. 
	\item We build a prototype to validate the proposed multiperson detection and vital-sign monitoring system. Experimental results demonstrate that the proposed system can detect up to 4 persons, and the RR and HR estimation errors with the proposed VMD algorithm are below 1 RPM (respiration per minute) and 5 BPM (beats per minute), respectively.
\end{itemize}

The remainder of this paper is organized as follows. First, we briefly overview the proposed multiperson detection and vital-sign monitoring system in Section II, which consists of the physical layer, the digital layer and the sensing layer. Subsequently, the basic theory of the SCT RISs is briefly introduced in Section III. In Section IV, the signal model of the STC RIS-based multiperson sensing is presented. Next, an improved VMD algorithm is proposed for simultaneous RR and HR estimation in Section V. Furthermore, a prototype is built to validate the proposed human detection and vital-sign monitoring system, and the experimental setup and result analysis are presented in Section VI. Finally, Section VII concludes this paper.

\section{System Overview}
As illustrated in Fig. \ref{overall}, the proposed multiperson detection and vital-sign monitoring system comprises three layers: the physical layer, the digital layer, and the sensing layer. Through these three modules, the system can consistently detect the presence of human targets while simultaneously performing RR and HR estimation for the identified individuals. The configuration and functionality of each layer are outlined as follows.

\subsection{Physical Layer}
The physical layer consists of a transceiver module and an STC RIS module. The transceiver module utilizes a universal software radio peripheral (USRP) device connected to two antennas for signal transmission and reception. Meanwhile, in the RIS module, a microprogrammed control unit (MCU) is employed to control the RIS hardware. With the STC manner, the RIS can manipulate the incident EM waves and customize wireless sensing channel in both spatial and frequency domains. By this means, human detection and vital-sign monitoring could be performed simultaneously by utilizing different harmonic beams for various tasks. Furthermore, to facilitate integrated human detection and vital-sign sensing, system-level communication is indispensable. Therefore, a host computer is integrated into the system as the ``brain", responsible for communication with the USRP and the MCU, as well as for echo signal processing.
\subsection{Digital Layer}
The digital layer is primarily utilized for human target detection. Specifically, the harmonic waves generated by the STC RIS are employed for beam scanning within the AoI, which is similar to \textit{Lidar}. In each scanning, the received echo signals with different harmonic frequencies could be disentangled in the frequency domain, and then utilized for human detection. Two indicators, namely the strength of the reflected signal and the respiratory sign, are adopted to deduce the presence of human targets in the currently scanned area. If a human target is detected, a harmonic beam will be assigned to this direction, and the echo signals of this individual will be collected for fine-grained vital-sign sensing. Meanwhile, the other harmonic beams not employed for vital-sign sensing will continue scanning the AoI for human detection.
\subsection{Sensing Layer}
\begin{figure}[h]
	\centering
	\includegraphics[width=1\columnwidth]{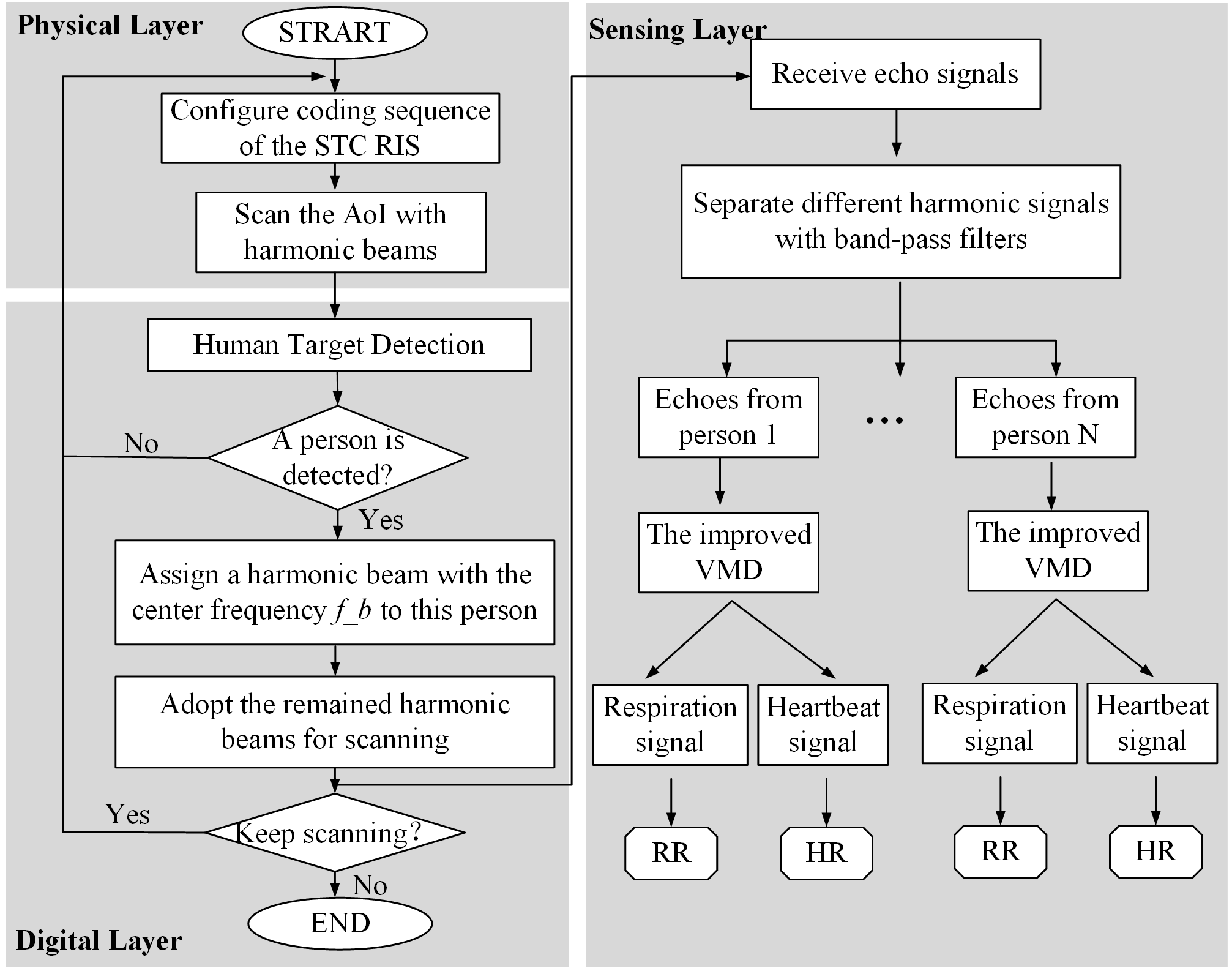}
	\caption{Signal processing pipeline of the proposed multiperson detection and vital-sign sensing system.}
	\label{fig2}
\end{figure}

In the sensing layer, as the detected human targets are illuminated by different harmonic beams, their echo signals can be separated in the frequency domain using a series of band-pass filters. Furthermore, the beamforming empowered by the STC RIS can enhance the SNR of the reflected signal from the target individual and reduce the interference from other persons and environmental clutters. Subsequently, an improved VMD algorithm is proposed to extract physiological signals from the reflected echoes. By decomposing the received signals into a series of narrow-band components, the sub-signals associated with respiration and heartbeat can be extracted from the complex received signals. Moreover, with the proposed filter-based signal reconstruction mechanism and the adaptive penalty factor, the proposed VMD algorithm can accurately extracting physiological signals with a small number of IMFs. The whole signal processing pipeline of the multiperson detection and vital-sign sensing system is summarized in Fig. \ref{fig2}.

\section{Basic Theory of The STC RISs}
\label{s3}
In this paper, to focus harmonic beams modulated by the STC RIS on different human bodies and thereby reduce interference from other background components, the meta-atom units of the STC RIS are controlled in both time and spatial domains. 
Specifically, when a tone signal illuminates the STC RIS, the near-field scattering pattern in the time domain is the superposition of radiation from all cells inside the RIS, which is formulated as \cite{zhang2018space}
%%%%%%%%%%%%%%%%%%%%%%%%%%%%%%%%%%%%%%%%%%%%%%%%%%%%%%%%%%%%%%%%%%%%%%%%%%%%%
\begin{equation}
    g_{i,j}(t) =  A_{m,n}\sin(2\pi f_ct+\phi_{m,n})\sum_{m=1}^{M}\sum_{n=1}^{N} w_{i,j,m,n} \Gamma_{m,n}(t) \label{STC:1}
\end{equation}
where the single tone illuminating the $(m,n)$th element of the RIS is denoted by $A_{m,n}\sin(2\pi f_ct+\phi_{m,n})$ with a known amplitude $A_{m,n}$, center frequency $f_c$, and phase $\phi$. $\Gamma_{m,n}(t)$ denotes the programmable reflection coefficient of the $(m, n)$th element in the RIS, and $g_{i,j}(t)$ represents the received time-domain signal at the $(i,j)$th element of the target region. Then, based on the near-field diffraction model, the transmission parameter $w_{i,j,m,n}$ is defined as 
\begin{equation}
    w_{i,j,m,n} = \frac{z}{\lambda}\left(\frac{1}{kr}+\frac{1}{j}\right)\frac{1}{r^2}\exp(jkr)\label{STC:2}
\end{equation} 
where $r$ is the distance between the $(i,j)$th element of the target region and the $(m,n)$th element of the RIS, $z$ is the distance between the RIS and the target region, and $k=2\pi / \lambda$ is the wave-number. 
%%%%%%%%%%%%%%%%%%%%%%%%%%%%%%%%%%%%%%%%%%%%%%%%%%%%%%%%%%%%%%%%%%%%%%%%%%%%%

Subsequently, the STC strategy is employed to modulate different frequency components of the scattering pattern to different orientations, as illustrated in Fig. \ref{overall}(b). In this manner, the sensing signals directed at different individuals are distinguishable in both the spatial and frequency domains.
To achieve this, a set of coding sequences is cyclically switched within a pre-designed time period, modulating the meta-atoms in the time domain. Then, the frequency-domain near-field scattering pattern $G_{i,j}(f)$ is expressed as

\begin{align}\label{STC:3}
    G_{i,j}(f) = &\delta(f-f_c) \circledast
    \\ \notag &\sum_{m=1}^{M}\sum_{n=1}^{N} A_{m,n}e^{j\phi_{m,n}}w_{i,j,m,n} FT\{\Gamma_{m,n}(t)\},
\end{align}
where $G_{i,j}(f)$ is the signal spectrum at the $(i,j)$th element of the target region. $FT\{\cdot\}$ represents the Fourier transform, and $\circledast$ is the convolution operation. The periodic time-modulated reflection coefficient $\Gamma_{m,n}(t)$ in Eq. (\ref{STC:3}) is defined as a linear combination of shifted pulse function, which is written as

\begin{equation}
	\Gamma_{m,n}(t) = \sum_{l=1}^{L} \Gamma_{m,n}^{l}U_{m,n}^l(t), 0 \leq t \leq T_0,\label{STC:4}
\end{equation}
where $U_{m,n}^l(t)$ is a periodic pulse function with modulation period $T_0$. In each period, we have 

\begin{equation}
   U_{m,n}^l(t) = \left \{	
   \begin{array}{rr}			
   	1,&(l-1)\tau \leq t \leq l\tau,  \\			
   	0, &otherwise,			
   \end{array}	
   \right.
\end{equation}
where $\tau=T_0/L$ is the pulse width of $U_{m,n}^l(t)$, $L$ is a positive integer representing the length of the time-coding sequence, $\Gamma_{m,n}^{l} = B_{m,n}^le^{j\beta_{m,n}^l}$ is the reflection coefficient of the $(m, n)$th unit at the center frequency during the interval $(l-1)\tau \leq t \leq l\tau$, and $B_{m,n}$ and $\beta_{m,n}$ represent the modulated amplitude and phase of the $(m, n)$th unit, respectively. For a 1-bit RIS, the reflection amplitude $B_{m,n}$ of each unit is uniform, whereas the phase $\beta_{m,n}$ is either 0° or 180°, according to the digits ``0" and ``1" in the coding sequence.
%%%%%%%%%%%%%%%%%%%%%%%%%%%%%%%%%%%%%%%%%%%%%%%%%%%%%%%%%%%

Next, by decomposing $U_{m,n}^l(t)$ into a Fourier series, $FT\{\Gamma_{m,n}(t)\}$ in Eq. (\ref{STC:3}) can be formulated as 
\begin{align}\label{STC:6}
    FT\{\Gamma_{m,n}(t)\} & = \sum_{l=1}^{L} \Gamma_{m,n}^{l} FT\{U_{m,n}^l(t)\}
    \\ \notag & = \sum_{l=1}^{L} \Gamma_{m,n}^{l} \sum_{k=-\infty}^{+\infty}C_{k,l}\delta(f-kf_{sub}),
\end{align}
 where
\begin{equation}\label{STC:5}
    C_{k,l}=\frac{1}{2L}\exp{\left(\frac{j(1-2l)k\pi}{L}\right)}Sa\left(\frac{k\pi}{2L}\right),
\end{equation}
and
\begin{equation}
 Sa(x) = \sin{x}/x.
\end{equation}

Finally, substituting Eq.(\ref{STC:3}) and Eq. (\ref{STC:5}) into Eq. (\ref{STC:6}), the near-filed scattering pattern $G_{i,j}(f)$ can be rewritten as
\begin{align}\label{TIME-CODING-FINAL}
     G_{i,j}(f) = &\sum_{k=-\infty}^{+\infty}\delta(f-(f_c+kf_{sub}))
     \\\notag&\sum_{l=1}^{L}\sum_{m=1}^{M}\sum_{n=1}^{N} A_{m,n}e^{j\phi_{m,n}}w_{i,j,m,n}
     C_{k,l}\Gamma_{m,n}^{l}
\end{align}

From Eq. (\ref{TIME-CODING-FINAL}), we can find that the power distributions and the spatial propagation of the harmonic beams can be simultaneously controlled by optimizing the STC coding $\Gamma_{m,n}^{l}$, which could be constructed as a three-dimensional data block $\Gamma\in C^{M\times N\times L}$.

\section{Signal Model of Passive Human Sensing with STC RISs}

In the proposed system, a tone signal $x_T(t) = \exp(2\pi f_ct)$ is transmitted with a directional antenna Tx and directed towards the RIS. With the directional antenna Tx, it can be assumed that there is no line-of-sight (LoS) between the Tx and the AoI. In other words, the transmitted signal $x_T(t)$ is first modulated by the RIS and then reflected to the AoI for sensing.

Denote a diagonal matrix $\textbf{H}_{RIS} \in \mathbb{C}^{MN \times MN}$ as the manipulating matrix of the RIS with $M \times N$ units. The diagonal elements of $\textbf{H}_{RIS}$ form a vector $\textbf{h}_{RIS} = [B_{1,1}e^{j\beta_{1,1}}, ..., B_{1,N}e^{j\beta_{1,N}},...,B_{M,N}e^{j\beta_{M,N}}]$. Then, the signal $\textbf{x}_{RIS}(t)$ reflected by the RIS can be denoted as
\begin{equation}
	\textbf{x}_{RIS}(t) =  \textbf{H}_{RIS}(t)\textbf{H}_{T,RIS}(t)\textbf{x}_T(t) + \textbf{z}_{RIS,T}(t),
\end{equation}
where $\textbf{H}_{T,RIS}(t) \in \mathbb{C}^{MN \times 1}$ denotes the channel between the Tx and the RIS, $\textbf{z}_{RIS,T}(t)$ denotes the noise between the Tx and the RIS, and $\textbf{x}_T(t)$ represents the transmitted symbol at the time $t$. With the STC strategy introduced in Section \ref{s3}, the incident signal $\textbf{x}_T(t)$ is modulated into a series of harmonic components with diverse frequencies. Therefore, the signal $\textbf{x}_{RIS}(t)$ can be reformulated as a combination of a series of harmonic waves:

\begin{equation}
\begin{split}
\textbf{x}_{RIS}(t) &= \sum_{i} \textbf{x}^h_i(t) \\&= \textbf{x}_T(t)\sum_{i} A_{i}(t)\exp(2\pi if_0t) + \textbf{z}_{RIS,T}(t),\\ &(i=0, \pm1, ..., \pm \infty)
\end{split}
\end{equation}
where $f_0 = \frac{1}{T_0}$ is the time-domain modulation frequency of the STC RIS, and $A_{i}(t)$ is the amplitude of the $i$th harmonic wave.

Then, when there exists a human target illuminated by the $i$th harmonic beam, the signal $\textbf{x}^h_i(t)$ will be modulated by the movements of the individual. The reflected echoes of this individual are subsequently received by the omnidirectional antenna Rx at the receiving terminal. Thus, the channel model $\textbf{H}_{RIS,P,R}$ between the RIS and the Rx can be represented as

\begin{equation}
	\textbf{H}_{RIS,P,R}(t) = \textbf{H}_{P,R}(t)\Gamma^p_{i}(t)\exp(2\pi f_Dt)\textbf{H}_{RIS,P}(t),
\end{equation}
where $\textbf{H}_{P,R}(t)$ represents the channel between the Rx and the person illuminated by the $i$th harmonic beam, $\textbf{H}_{RIS,R}(t)$ represents the channel between the RIS and the person, $\Gamma^p_{i}(t) = B^p_i(t)e^{j\beta^p_{i}}(t)$ denotes the complex-valued reflected coefficient of the person at the time $t$, and $f_D$ stands for the Doppler frequency shift induced by the movement of this person.
%where $\textbf{H}_{P,R}(t)$ is the channel between the person and the Rx, $A'_i(t)$ is the amplitude of the reflected signal, and $f_D$ is the Doppler frequency caused by the chest movements from vital signs, $\textbf{H}_{RIS,P}(t)$ is the channel between the RIS and the person, and $\textbf{z}^h_{i,R}(t)$ is the noise.

As a result, considering there are two main propagation paths between the RIS and the Rx (i.e., the LoS path and the ``RIS-person-Rx" path), the received signal $\textbf{x}_R(t)$ composed of a series of harmonic components could be formulated as

\begin{equation}
	\textbf{x}_R(t) = \sum_{i} \textbf{H}_{i}(t) \textbf{H}_{RIS}\textbf{H}_{T,RIS}(t)\textbf{x}_T(t) + \textbf{z}_R(t),
\end{equation}
where $\textbf{z}_R(t)$ denotes the noise between the Tx and the Rx, and $\textbf{H}_{i}(t)$ can be formulated as

\begin{equation}
	\textbf{H}_{i}(t) = \left \{	
	\begin{array}{ll}			
		\textbf{H}_{RIS,P,R}(t) +\textbf{H}_{RIS,R}(t), & \text{there is a person} ,\\			
		\textbf{H}_{RIS,R}(t), & \text{there is no person},		\\	
	\end{array}	
	\right.		
\end{equation}

Since the frequencies of these harmonic waves are orthogonal, the reflected signals can be easily separated in the frequency domain using a series of band-pass filters, and subsequently utilized for human detection and vital-sign monitoring.

For the human detection task, we propose two indicators to identify whether there is a person at the illuminated direction. Specifically, we first prescan the whole AoI when there is no person, and obtain an intensity vector $\textbf{I}$ = [$I^N_{1}$, $I^N_{2}$, ..., $I^N_{D}$] of the received echo signals from $D$ directions. Then, a threshold $\mu$ of signal intensity is preset based on experiential knowledge. As a result, for the direction $d \in [1, D]$, the difference of the signal strength $\Delta I_d$ can be formulated as
\begin{equation}
	 \Delta I_d = I_d - I^N_{d},
\end{equation}
where $I_d$ is the strength of the echo signal from the direction $d$. If $\Delta I_d$ is greater than the preset $\mu$, the system infers that there may be a human individual on this direction. Furthermore, to recheck the target is a human individual, respiration detection is also employed. It is noted that some preliminary experiments have demonstrated that extracting the respiration signal is more accessible than obtaining the heartbeat signal \cite{9423591, 9217947, 10.1145/3377165}. Therefore, we adopt respiration signal as another indicator of detecting human individuals. In other words, if the respiration signal is extracted, the candidate is ultimately identified as a human individual. 

\section{Improved VMD for vital-sign sensing}
%Vital signs signals, i.e., the respiration signal $s_r(t)$ and the heartbeat signal $s_h(t)$, are narrow-band with center frequencies $w_r$ and $w_h$, respectively.

The VMD algorithm \cite{dragomiretskiy2013variational} has been successfully applied in various fields, including disease surveillance \cite{8474360}, seismic signal processing \cite{9783094}, and fault diagnosis \cite{10089181}. However, due to the fact that the tiny chest movements caused by heartbeats generally have little effect on the wireless signal propagation, the performance of using these baseline methods to extract heartbeat signal is limited. Furthermore, the heartbeat signal may be regarded as higher harmonics of respiration signal, causing difficulty in separating the two physiological signals. To deal with these issues, an improved VMD algorithm is proposed for physiological signal extraction.

As a non-recursive decomposition method, VMD can adaptively decompose a complicated multi-component signal $s(t)$ into a series of amplitude-modulated-frequency-modulated (AM-FM) signals. The AM-FM signals are the so-called \textit{IMFs}, which are written as:

\begin{equation}
	u_i(t) = A_i(t)\exp(\phi_i(t)),
\end{equation} 
where both $w_i(t)$ $:=$ $\phi_i'$ and the envelope $A_i(t)$ are non-negative. Furthermore, $A_i(t)$ and $w_i(t)$ vary much slower than $\phi_i$. Then, all the IMFs together with the residual reconstruct the original signal $s(t)$, which can be denoted as:

\begin{equation}
	s(t) = \sum_{i=1}^{I} u_i(t) + u_r(t),
\end{equation}
where $u_i(t)$ is the $i$th IMF, $I$ is the number of IMFs, and $u_r(t)$ is the residual. Some of these IMFs belong to the respiration signal $s_r(t)$, while others belong to the heartbeat signal $s_h(t)$. The residual could be treated as the noise irrelevant to the vital-sign monitoring task. 

To extract a series of IMFs compacted around their center frequencies, the bandwidth of these IMFs needs to be minimized as follows:

\begin{equation}
	\mathcal{J} = ||\partial_t(\delta(t) + j \frac{1}{\pi t}) \circledast s(t)]e^{-jwt}||_2^2,
\end{equation}
where $\mathcal{J}$ is the bandwidth of the IMF. Then, the resulting constrained variational problem is the following:

\begin{equation}
	\label{e1}
\begin{split}
   \min_{\{u_i\},\{w_i\}}\{\sum_{i}||\partial_t[(\delta(t)+\frac{1}{\pi t}) \circledast u_i(t)]e^{-jw_it}||_2^2\},
   \\s.t. \sum_{i}u_i(t)=s(t)
\end{split}
\end{equation}
where $\{u_i\} := \{u_1, u_2, ..., u_I\}$ and $\{w_i\} := \{w_1, w_2, ..., w_I\}$ denote the sets of all IMFs and their center frequencies, respectively.

Then, to enforce the constraints strictly, augmented Lagrangian is adopted to solve the quadratic optimization problem in Eq. (\ref{e1}), which is depicted as

\begin{equation}
\label{e2}
\begin{split}
	\mathcal{L}(\{u_i\},\{w_i\},\lambda) := \alpha\sum_{i}||\partial_t[(\delta(t) + \frac{1}{\pi t}) \circledast u_i(t)]e^{-jw_it}||_2^2\\ + ||s(t)-\sum_{i}u_i(t)||_2^2 + <\lambda(t), s(t) - \sum_{i}u_i(t)>,
\end{split}
\end{equation}
where $\lambda$ denotes Lagrangian multiplier, and $\alpha$ is the penalty factor. A larger $\alpha$ could result in the IMFs with narrower band. To solve the minimization problem in Eq. (\ref{e2}), an iterative sub-optimizations named Alternate Direction Method of Multipliers (ADMM) \cite{dragomiretskiy2013variational} is adopted. 

Furthermore, to avoid the spectrum overlap between $s_r(t)$ and $s_h(t)$, we propose a filter-based signal reconstruction mechanism, and adopt two filters on the IMFs during the ADMM optimization process. Specifically, we assume that the first $M$th IMFs belong to $s_r(t)$, while the other IMFs belong to $s_h(t)$. Then, in each iteration of the ADMM optimization, we filter the high-frequency components in the first $M$th IMFs with a low-pass filter, and also remove the noise in the remained IMFs with a band-pass filter. In this case, the original echo signal $s(t)$ could be reconstructed as:

\begin{equation}
	\label{e4}
	s(t) = \sum_{m=1}^{M} u_m(t) \circledast f_l(t) + \sum_{n=1}^{I-M} u_n(t) \circledast f_b(t) + u_r(t),
\end{equation}
where $u_m(t)$ is the $m$th candidate IMF belonging to $s_r(t)$, $u_n(t)$ is the $n$th candidate IMF belonging to $s_h(t)$, $f_l(t)$ is the low-pass filter, and $f_b(t)$ is the band-pass filter.

Furthermore, a frequency-correlated penalty factor $\alpha$ is proposed in the improved VMD algorithm for adaptively extracting IMFs with narrower bands. Specifically, we propose an index-based penalty factor for each IMF. For the $i$th IMF ($i \leq I$), its penalty factor can be denoted as 

\begin{equation}
	\alpha_i = \alpha_{int}\exp(-\zeta||w_i-w_r||_2^2)
\end{equation}
where $\alpha_{int}$ denotes the initial penalty factor, and $\zeta$ is the scale factor. $w_r$ denotes the reference frequency, which is preset based on the observation that the respiration frequency and the heartbeat frequency are generally around 0.25 Hz and 1.35 Hz, respectively. With the adaptive $\alpha$, when the center frequency of the $i$th IMF is approaching its reference $w_r$, a greater penalty will be assigned to this IMF during the ADMM optimization process, resulting in a narrower-band signal and filtering the out-of-band noise.

Overall, the quadratic optimization problem in Eq. (\ref{e2}) could be reformulated as:

\begin{equation}
\label{e3}
\begin{split}
	\mathcal{L}(\{u_i\},\{w_i\},\lambda) := \sum_{i=1}^{M}||\alpha_i \partial_t[(\delta(t) + \frac{1}{\pi t}) \circledast f_l(t) \circledast u_i(t)]e^{-jw_it}||_2^2\\
	+ \sum_{i=M+1}^{I}||\alpha_i \partial_t[(\delta(t) + \frac{1}{\pi t}) \circledast f_b(t) \circledast u_i(t)]e^{-jw_it}||_2^2\\ + ||s(t)-\sum_{i}^If_i(t) \circledast u_i(t)||_2^2 + <\lambda(t), s(t) - \sum_{i}^If_i(t) \circledast u_i(t)>,\\
	s.t. \sum_{i=1}^{M} f_l(t) \circledast u_i(t) + \sum_{i=M+1}^{I} f_b(t) \circledast u_i(t)=s(t),
\end{split}
\end{equation}
where $f_i(t)$ denotes the filter implemented on the $i$th IMF, which could be a low-pass or a band-pass filter.

Then, based on Plancherel-Parseval Fourier isometry property under the L2 norm and changing $w \rightarrow w - w_i$, Eq. (\ref{e3}) could be rewritten as Eq. (\ref{e111}):

\begin{figure*}[h]
\centering
\begin{equation}\label{e111}
\begin{aligned}
	\mathcal{L}(\{\hat{u}_i\},\{w_i\},\lambda) :=& \sum_{i=1}^{M}||\alpha_i j(w-w_i)(1+sgn(w))\hat{f}_l(w)\hat{u}_i(w)||_2^2 + \sum_{i=M+1}^{I}||\alpha_i j(w-w_i)(1+sgn(w))\hat{f}_b(w)\hat{u}_i(w)||_2^2\\&+ ||\hat{s}(w)-\sum_{i}\hat{f}_i(w)\hat{u}_i(w)||_2^2 + <\hat{\lambda}(w), \hat{s}(w) - \sum_{i}\hat{f}_i(w)\hat{u}_i(w)>,\\
	&s.t. \sum_{i=1}^{M} \hat{f}_l(w)\hat{u}_i(w) + \sum_{i=M+1}^{I} \hat{f}_b(t)\hat{u}_i(t)=\hat{s}(w),
\end{aligned}	
\end{equation}
\end{figure*}
%where $\hat{u}_i(w)$, $\hat{s}_i(w)$, $\hat{f}_l(w)$, $\hat{f}_b(w)$ and $f_i(w)$ are the Fourier transform of $u_i(t)$, $s_i(t)$, $f_l(t)$, $f_b(t)$, and $f_i(t)$ respectively.
%\\ \hspace*{\fill} \\

Finally, with the ADMM algorithm, the $i$th IMF $\hat{u}_i(w)$ in the $(n+1)$th iteration could be calculated as Eq. (\ref{e222}):

\begin{equation}\label{e222}
	\begin{split}		
		\hat{u}_i^{n+1}(w) = \left \{	
		\begin{array}{rr}			
			\frac{\hat{s}(w)-\sum_{k \neq i}\hat{f}_l \hat{u}_k^n(w) + \frac{\hat{\lambda}(w)}{2}}{1 + 2\alpha_i(w-w_i^n)},                    & 1 \leq i \leq M \\			
			\frac{\hat{s}(w)-\sum_{k \neq i} \hat{f}_b \hat{u}_k^n(w) + \frac{\hat{\lambda}(w)}{2}}{1 + 2\alpha_i(w-w_i^n)},     & M \textless i \leq I			
		\end{array}	
		\right.				
	\end{split}	
\end{equation}

Meanwhile, the center frequency $w_i^{n+1}$ of $\hat{u}_i^{n+1}(w)$ could be calculated as Eq. (\ref{e333}):

\begin{equation}\label{e333}
	w_i^{(n+1)} = \frac{\int_{0}^{\infty} w|\hat{u}_i^{n+1}(w)|^2 dw}{\int_{0}^{\infty} |\hat{u}_i^{n+1}(w)|^2 dw}
\end{equation}

\begin{algorithm}[h]
	\caption{The Improved VMD Algorithm.}
	\label{alg:algorithm1}
	\KwIn{The received echo signal $s(t)$, the total number $I$ of IMFs, the number $M$ of the candidate IMFs belonging to $s_r(t)$, Lagrange multiplier $\hat{\lambda}_0$, maximum iterations $Iter_{max}$, mode convergence absolute tolerance $\varepsilon_a$, mode convergence relative tolerance $\varepsilon_r$, and penalty factor $\alpha_i$ ($i$ = 1, 2, ..., $I$).}
	\KwOut{The respiration signal $s_r(t)$ and the heartbeat signal $s_h(t)$.}  
	
	\textbf{Initialize} Index $n=0$
	
    \While{\textnormal{$n \leq Iter_{max}$} and \textnormal{
    $\sum_{i}^{I}||u_i^{n+1}(t)-u_i^{n}(t)||_2^2/||u_i^n(t)||_2^2 \leq \varepsilon_a$} and \textnormal{$\sum_{i}^{I}||u_i^{n+1}(t)-u_i^{n}(t)||_2^2 \leq \varepsilon_r$}}{
    \textbf{Initialize} Index $i$ = 1
    
    \While{\textnormal{$i \leq I$}}{
    \If{\textnormal{$i \leq M$}}
	{\textbf{Update} $u^{n+1}_i(w)$ with:
	
	$\hat{u}^{n+1}_i(w) \leftarrow \frac{\hat{s}(w)-\sum_{k \neq i}\hat{f}_l \hat{u}_k^n(w) + \frac{\hat{\lambda}(w)}{2}}{1 + 2\alpha_i(w-w_i^n)}$}
	\ElseIf{\textnormal{$M+1 \leq i \leq I$}}{\textbf{Update} $u_i^{n+1}(w)$ with:
		
		$\hat{u}_i^{n+1}(w) \leftarrow \frac{\hat{s}(w)-\sum_{k \neq i} \hat{f}_b \hat{u}_k^n(w) + \frac{\hat{\lambda}(w)}{2}}{1 + 2\alpha_i(w-w_i^n)}$}
	\textbf{Update} $w_i^{(n+1)}$ with:
	$w_i^{(n+1)} \leftarrow \frac{\int_{0}^{\infty} w|\hat{u}_i^{n+1}(w)|^2 dw}{\int_{0}^{\infty} |\hat{u}_i^{n+1}(w)|^2 dw}$}

	\textbf{Update} $\hat{\lambda}^{n+1}$ with:
		
	$\hat{\lambda}^{n+1} \leftarrow \hat{\lambda}^{n} + \epsilon[\hat{s}(w)-\sum_{i=1}^{I}\hat{u}_i^{n+1}(w)]$
	}

\end{algorithm}

\section{Experiment Implementation and Results Analysis}
\subsection{Implementation}
\subsubsection{Experimental Setup}

\begin{figure*}[h]
	\centering
	\includegraphics[width=1.8\columnwidth]{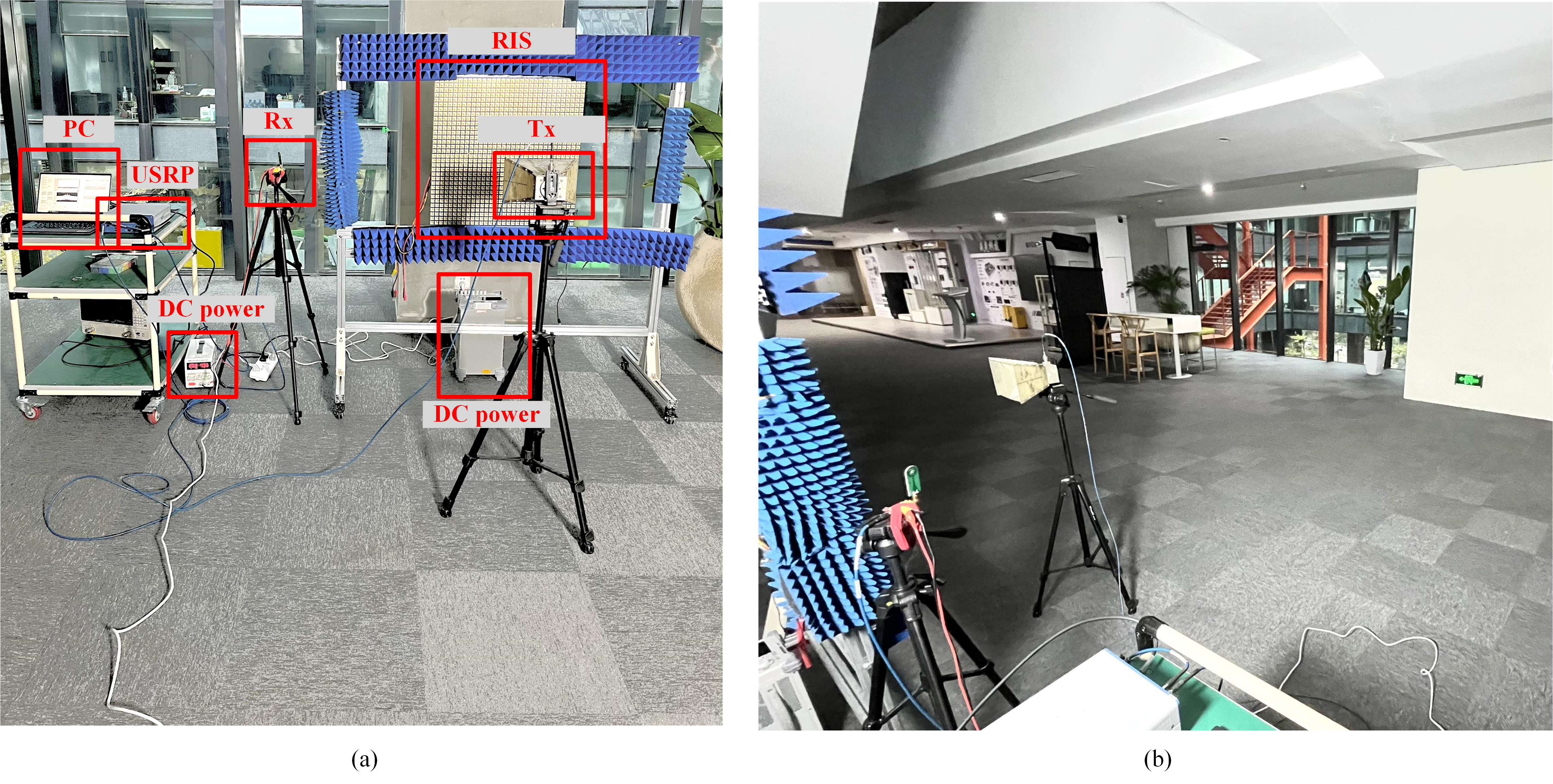}
	\caption{Experimental setup. (a) Experimental deployment, and (b) sensing area. The system is deployed in the corridor, containing multiple chairs, desks, and occasionally someone would pass by.}
	\label{setup}
\end{figure*}

The experimental setup is shown in Fig. \ref{setup}. The system is deployed in the corridor, containing multiple chairs, desks, and occasionally someone would pass by. The USRP device \textit{USRP-2974} is utilized as the sensing signal generator. Meanwhile, a high-gain horn antenna is in front of the RIS to transmit signals, and an omnidirectional antenna is placed on one side of the RIS for receiving signals. The two antennas are fixed on the tripods at a height of 1.2 m, respectively.

\begin{figure}
	\centering
	\includegraphics[width=\columnwidth]{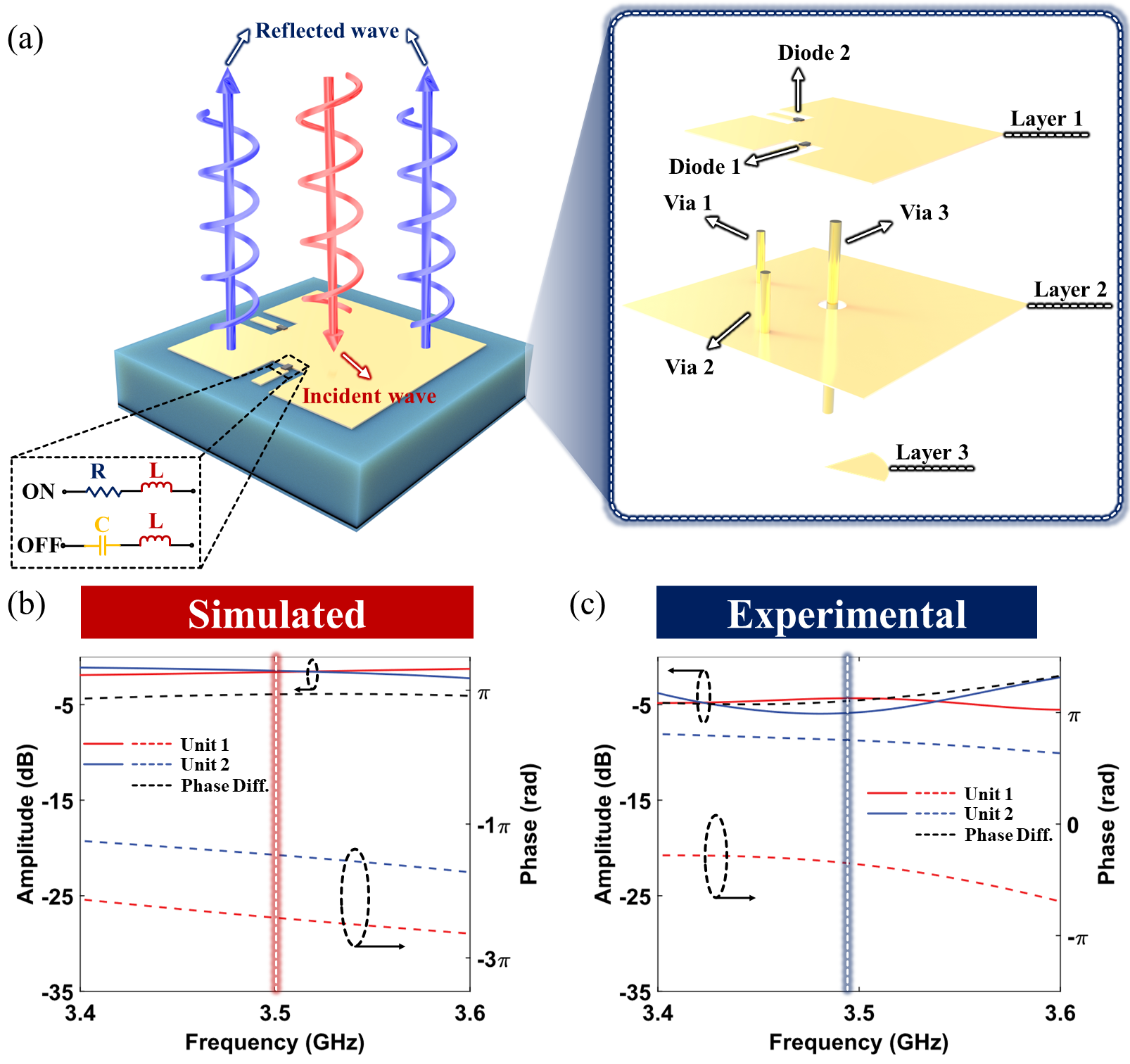}
	\caption{The employed 1-bit RIS. (a) The meta-atom unit, (b) the simulated magnitude-frequency and phase-frequency response of the designed meta-atom, and (c) the experimental magnitude-frequency and phase-frequency response of the designed meta-atom. }
	\label{atom}
\end{figure}

Besides, we integrate a 1-bit RIS \cite{jiang2023simultaneously} operating at $\sim$3.5 GHz into the proposed passive multi-person sensing system. The employed RIS consists of 32$\times$32 meta-atom units. The digital meta-atom includes two PIN diodes, each of which is integrated to electrically and dynamically control the reflected EM response. From the magnitude-frequency and phase-frequency response of the designed meta-atom in Fig. \ref{atom}, it could be found that there is nearly no amplitude difference in the reflected EM waves between the ``0" and ``1" states at around 3.5 GHz, while the corresponding phase difference is approximately 180 degrees. As a result, the meta-atom can be effectively utilized in the design of a phase-based digital RIS. 

\subsubsection{Ground Truth}
To acquire accurate measurements of RR and HR as the ground truth, we employ two wearable sensors, namely \textit{HKH-11C} for respiration monitoring and \textit{HKD-10C} for heartbeat monitoring. The \textit{HKH-11C} device estimates RR by utilizing piezoelectric materials to detect abdominal pressure caused by human breathing, while \textit{HKD-10C} obtains HR by capturing ECG signals.

\subsubsection{Approaches for Comparison}
To demonstrate the feasibility and efficiency of the proposed vital-sign monitoring system, we compare the proposed method with other state-of-the-art (SOTA) approaches in \cite{li2023metaphys}, \cite{10247237} and \cite{xia2023metabreath}. 
\subsubsection{Metrics}
In this paper, RR and HR are measured by RPM and BPM, respectively. Then, the sensing error is defined as the absolute value of the difference between the estimated RPM (or BPM) $R_e$ and its corresponding ground truth $R_g$, i.e., $|R_e - R_g|$. 
\subsubsection{Software}
In this system, the software \textit{Labview 2019} from National Instruments (NI) is adopted to control the whole system. Specifically, a Labview project is created in which the baseband sensing signal is first generated with the \textit{Labview} code and subsequently sent to the air by the USRP. Then, the signal processing module for target detection and vital-sign sensing is implemented with Matlab script and then integrated into the \textit{Labview} project. Meanwhile, serial port communication between the host computer and the MCU is also realized with \textit{Labview}. By this means, the coding sequence of the STC RIS in the next scanning would change under the instruction sent by the host computer, which is determined by the current signal processing result. 

%%%%%%%%%%%%%%%%%%%%%%%%%%%%%%%%%%%%%%%%%%%%%%%%%%%%%%%%
\subsection{Generation of the STC Coding Sequences}
\begin{figure*}
	\centering
	\includegraphics[width=1.95\columnwidth]{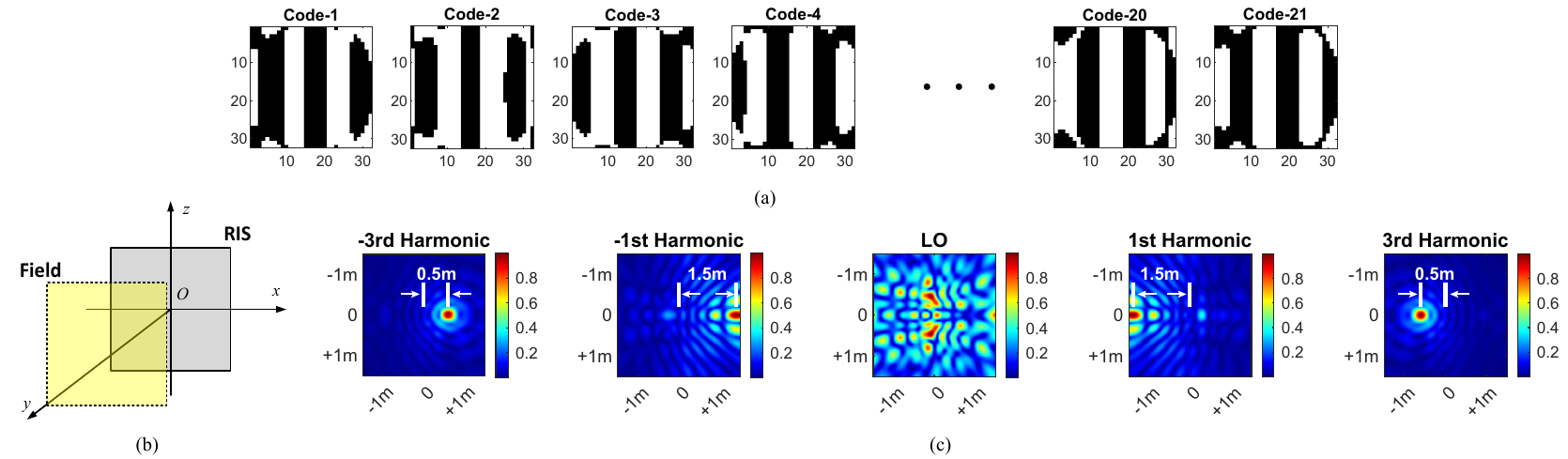}
	\caption{STC coding generation. (a) The 21 coding sequences optimized with the BPSO algorithm, (b) the reference coordinate system, and (c) the near-field patterns of the $-3rd, -1st, +1st, +3rd$ harmonic beams, which focus their energy on the positions (-1.5 m, +1 m, 0 m), (-0.5 m, +1 m, 0 m), (+0.5 m, +1 m, 0 m) and (+1.5 m, +1 m, 0 m), respectively. LO represents the local oscillator, or in other words, the zero harmonic component.}
	\label{timecoding}
\end{figure*}

Based on the numeric model derived in Eq.(\ref{TIME-CODING-FINAL}), the STC coding patterns of the RIS can be generated using the Binary Particle Swarm Optimization (BPSO) algorithm \cite{zhang2018space}, which enables the generated harmonic waves to concentrate on specific positions. For example, to direct the $-3rd, -1st, +1st, +3rd$ harmonic beams to the positions (-1.5 m, +1 m, 0 m), (-0.5 m, +1 m, 0 m), (+0.5 m, +1 m, 0 m) and (+1.5 m, +1 m, 0 m), respectively, a set of STC coding sequences is optimized using the BPSO algorithm, as depicted in Fig.\ref{timecoding}(a). Subsequently, the 21 coding patterns are injected into the RIS periodically at a frequency of $f_0$. As a result, the near-field patterns of the $-3rd, -1st, +1st, +3rd$ harmonic beams are customized through the time-domain modulation, as illustrated in Fig.\ref{timecoding}(b). It can be observed that the energy of these harmonics has been focused at the specified positions, demonstrating the efficiency of the generated STC coding sequences.

%%%%%%%%%%%%%%%%%%%%%%%%%%%%%%%%%%%%%%%%%%%%%%%%%%%%%%%%

\subsection{Multiperson Detection and Vital-Sign Monitoring}

\subsubsection{Experimental Results in Multiperson Scenarios}

\begin{figure*}
	\centering
	\includegraphics[width=2\columnwidth]{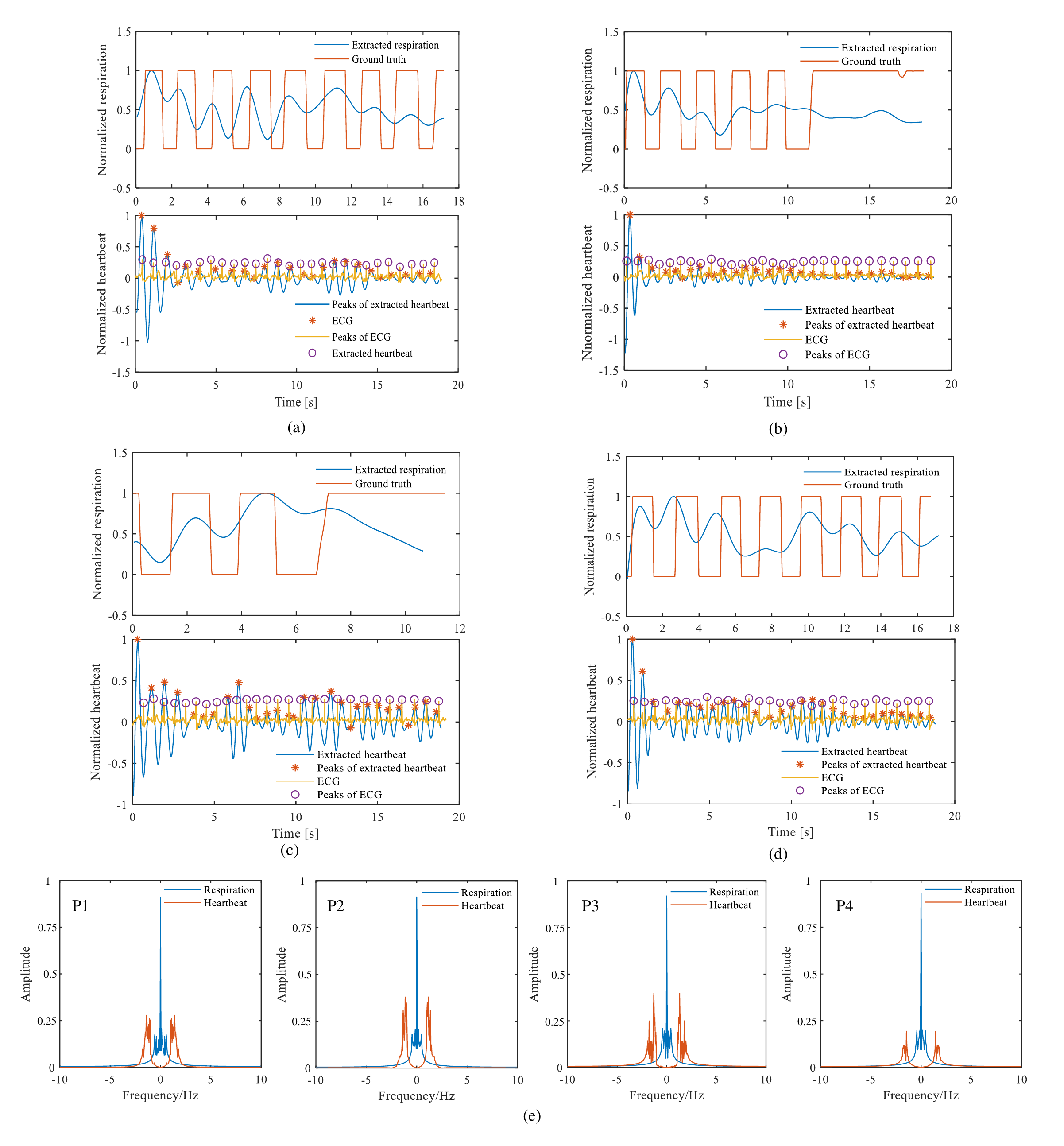}
	\caption{Multiperson vital-sign sensing results. (a)-(d) The time-domain respiration signal and heartbeat signal of the four persons. (e) The frequency-domain respiration signal and heartbeat signal of the four persons.}
	\label{1234}
\end{figure*}

The proposed system demonstrates the capability to simultaneously monitor the vital signs of up to four individuals with relatively low estimation errors. Fig. \ref{1234} illustrates the experimental results of multiperson vital-sign monitoring. Specifically, Fig. \ref{1234}(a)-(d) show the time-domain respiration signal and heartbeat signal of the four individuals, respectively. Fig. \ref{1234}(e) presents the corresponding frequency-domain spectrum of the physiological signals from the four persons. It can be seen that the physiological signals extracted by the proposed system exhibit similar periodicity to the ground-truth signals. Furthermore, as illustrated in Fig. \ref{1234}(d), the proposed algorithm can accurately detect breath-holding of the human subject, indicating the effectiveness of the proposed system in vital-sign sensing.

\begin{table}[h]
	\centering

	\caption{Human detection and vital-sign sensing results with different numbers of individuals to be perceived}
	\renewcommand\arraystretch{1.5}
	\begin{tabular}{l|lll}
		\hline
		\hline
\begin{tabular}[c]{@{}l@{}}\textbf{No. of} \\ \textbf{Persons}\end{tabular} & \begin{tabular}[c]{@{}l@{}}\textbf{No. of} \\ \textbf{Detected Persons}\end{tabular} & \begin{tabular}[c]{@{}l@{}}\textbf{Aver. error}\\  \textbf{of RR}\end{tabular} & \begin{tabular}[c]{@{}l@{}}\textbf{Aver. error}\\  \textbf{of HR}\end{tabular} \\
		\hline
		\textbf{1}              & 1                      & 0.5 RPM              & 3.7 BPM                 \\
		\hline
		\textbf{2}              & 2                      & 0.5 RPM               & 3.3 BPM                  \\
		\hline
		\textbf{3}              & 3                      & 1.5 RPM              & 5.8 BPM                  \\
		\hline
		\textbf{4}              & 4                      & 1.0 RPM                & 7.2 BPM     \\ 
		\hline   
		\hline        
	\end{tabular}
	\label{tq1}
\end{table}

\subsubsection{Performance Summary}

Table \ref{tq1} presents the estimation errors for RR and HR with varying numbers of individuals.  Specifically, experimental results indicate an average error of 1 RPM and 7.2 BMP for monitoring the vital signs of four persons. By assigning a harmonic beam to each detected person for sensing purposes, the spatial freedom achieved by the STC RIS significantly reduces mutual interference between echoes from different individuals, enhancing the accuracy of human detection and vital-sign sensing. Moreover, by flexibly manipulating the frequencies and spatial propagation of harmonic beams, the system can initiate or cease obtaining someone's physiological signals and estimate vital signs at any time. In comparison with the system proposed in \cite{xia2023metabreath}, our system can monitor both breathing and heartbeat of an individual, with each person monitored independently using a specific harmonic beam. Furthermore, compared to the system \cite{li2023metaphys} utilizing an SDC RIS for time-division multiperson sensing, our system can generate multiple frequency-orthogonal beams for simultaneous multiperson 
detection and vital signs 
sensing.                                                                                                            
\subsubsection{Algorithm Comparison}
\begin{figure}
	\centering
	\includegraphics[width=\columnwidth]{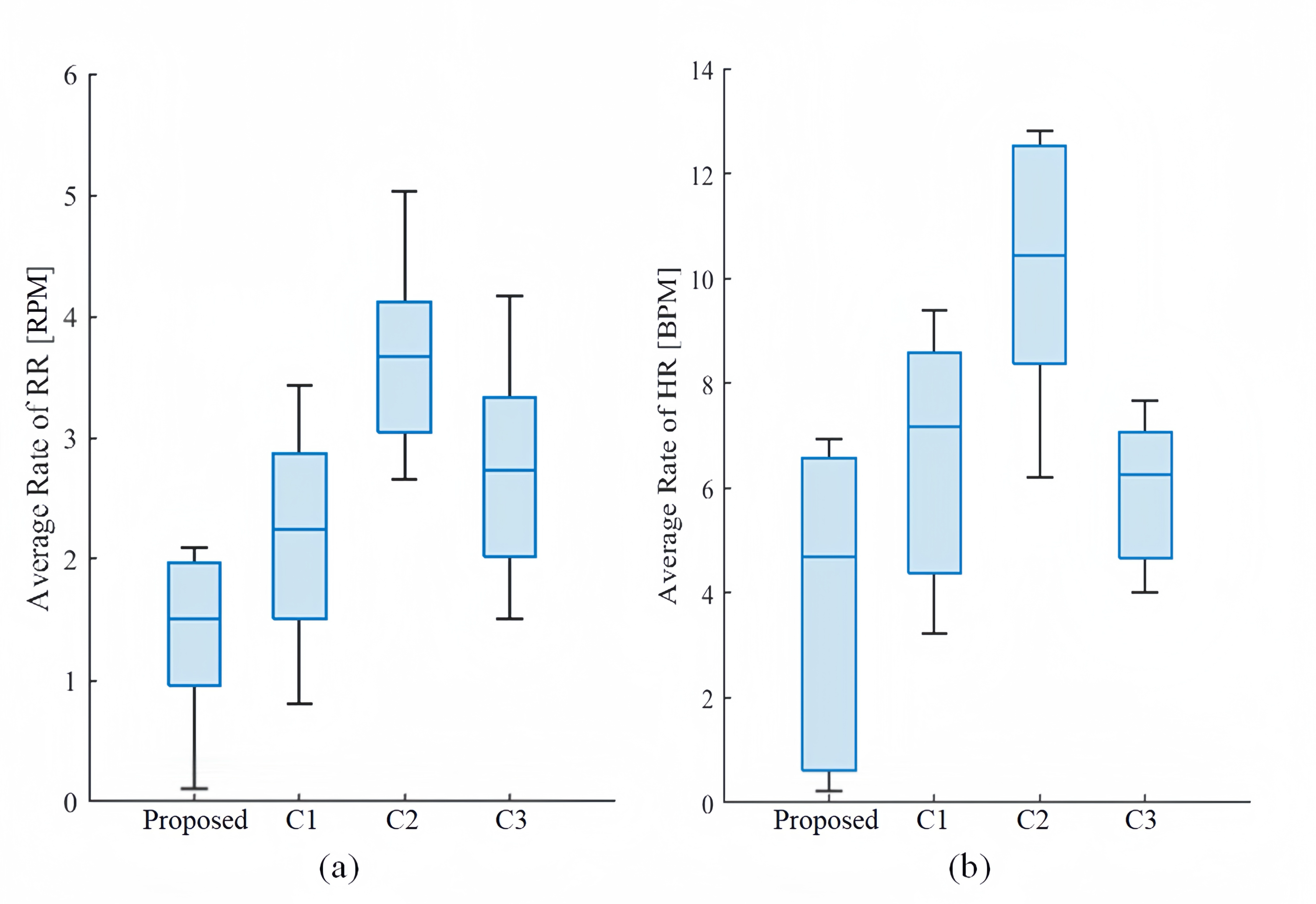}
	\caption{Experimental results of RR and HR estimation with the proposed method and other SOTA approaches. C1, C2, C3 denote the approaches in \cite{li2023metaphys}, \cite{xia2023metabreath}, and \cite{10247237}, respectively. }
	\label{compare}
\end{figure}

Fig. \ref{compare} presents the RR and HR estimation results of the proposed system compared to other SOTA systems. The proposed algorithm has demonstrated superior performance among the SOTA approaches for vital-sign monitoring. Specifically, for the RR estimation task, the average error of the proposed algorithm is approximately 1.6 BPM, significantly lower than that of the other counterparts. Moreover, the proposed improved VMD, with an estimation error for heartbeat sensing approximately 2 BPM lower than the algorithm in \cite{10247237}, contributes to a more substantial performance improvement in the HR estimation task.

\begin{figure}[h]
	\centering
	\includegraphics[width=\columnwidth]{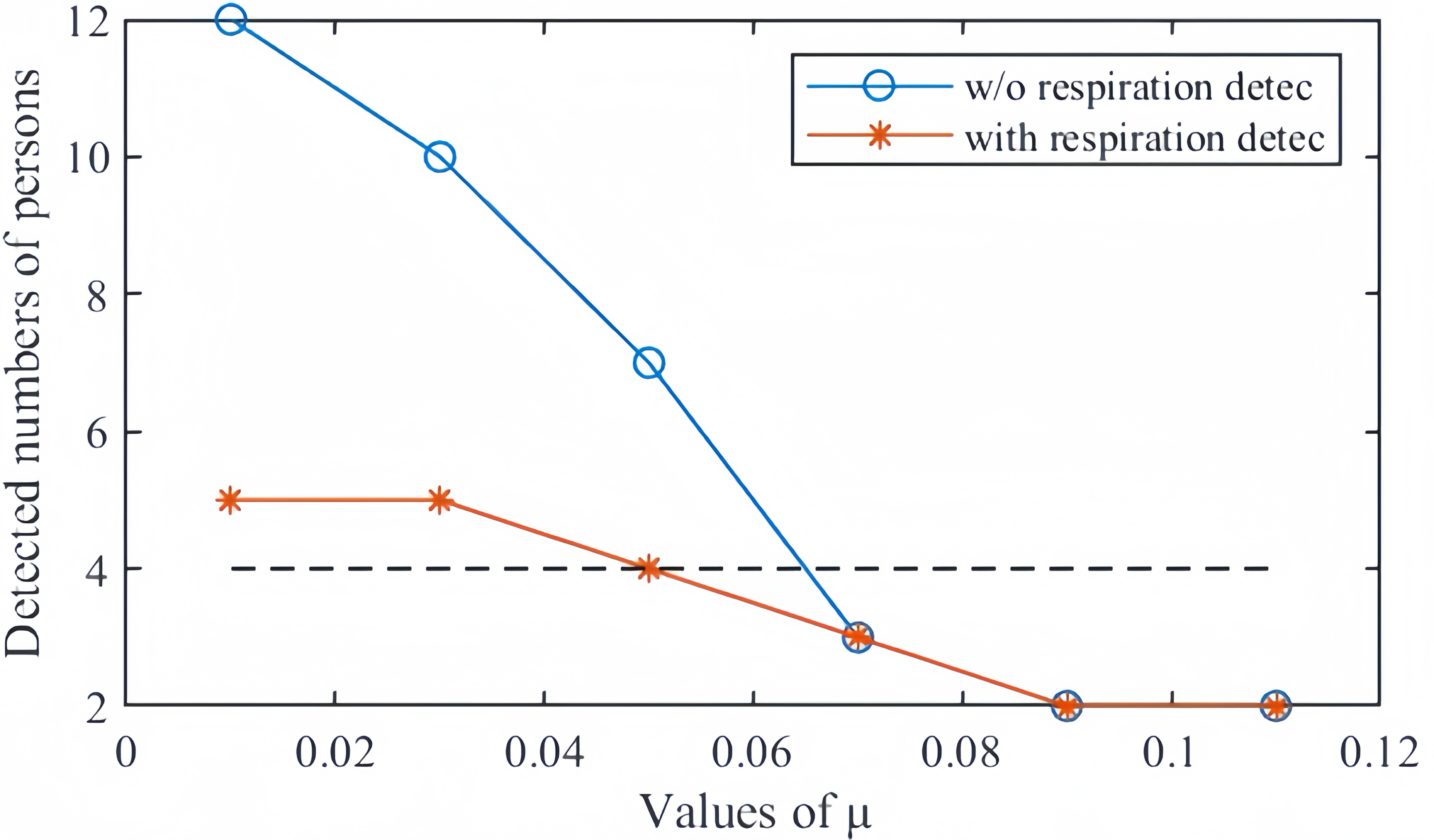}
	\caption{Human detection results using different indicators when there are four human individuals to be detected.}
	\label{mu}
\end{figure}

\subsection{Parameter Analysis on Multiperson Detection}
To demonstrate the efficiency of the proposed multiperson detection algorithm, we further analyze the impact of the adopted indicators (i.e., echo signal strength and the respiration signal) on human detection performance of the system. In detail, Fig. \ref{mu} shows the human detection results of the system using different indicators when there are four persons to be detected. It can be observed that the false alarm rate of target detection is high when only the signal strength indicator is utilized. However, with the assistance of the respiration indicator, the robustness of human detection is significantly improved. Furthermore, Fig. \ref{mu} also illustrates the multiperson detection results with different preset value of $\mu$ when the human targets are within 1 to 2 meters in front of the RIS. It can be seen that the value of $\mu$ has a significant effect on human detection performance. Therefore, we set the value of $\mu$ to 0.05 in the experiments based on empirical knowledge.
\subsection{Parameter Analysis on the Improved VMD}

\begin{table}[h]
	\centering
	\renewcommand\arraystretch{1.5}

	\caption{Average Estimation Errors of RR and HR with Different VMD Algorithms}
	\begin{tabular}{l|ll}
		\hline
		\hline
		& \textbf{Aver. Error of RR} & \textbf{Aver. Error of HR} \\
		\hline
		\textbf{Baseline VMD }                           & 2.7 RPM           & 13.0 BPM          \\
		\hline
		\textbf{IVMD w/o adpative} $\alpha$ & 2.0 RPM           & 8.8 BPM           \\
		\hline
		\textbf{IVMD w/o filters}                        & 1.6 RPM           & 6.9 BPM           \\
		\hline
		\textbf{IVMD}                                    & 1.1 RPM           & 4.7 BPM      \\   
		\hline 
		\hline
	\end{tabular}
	\label{t2}
\end{table}

\subsubsection{The Adaptive Penalty Factor $\alpha$}
\begin{figure}[h]
	\centering
	\includegraphics[width=\columnwidth]{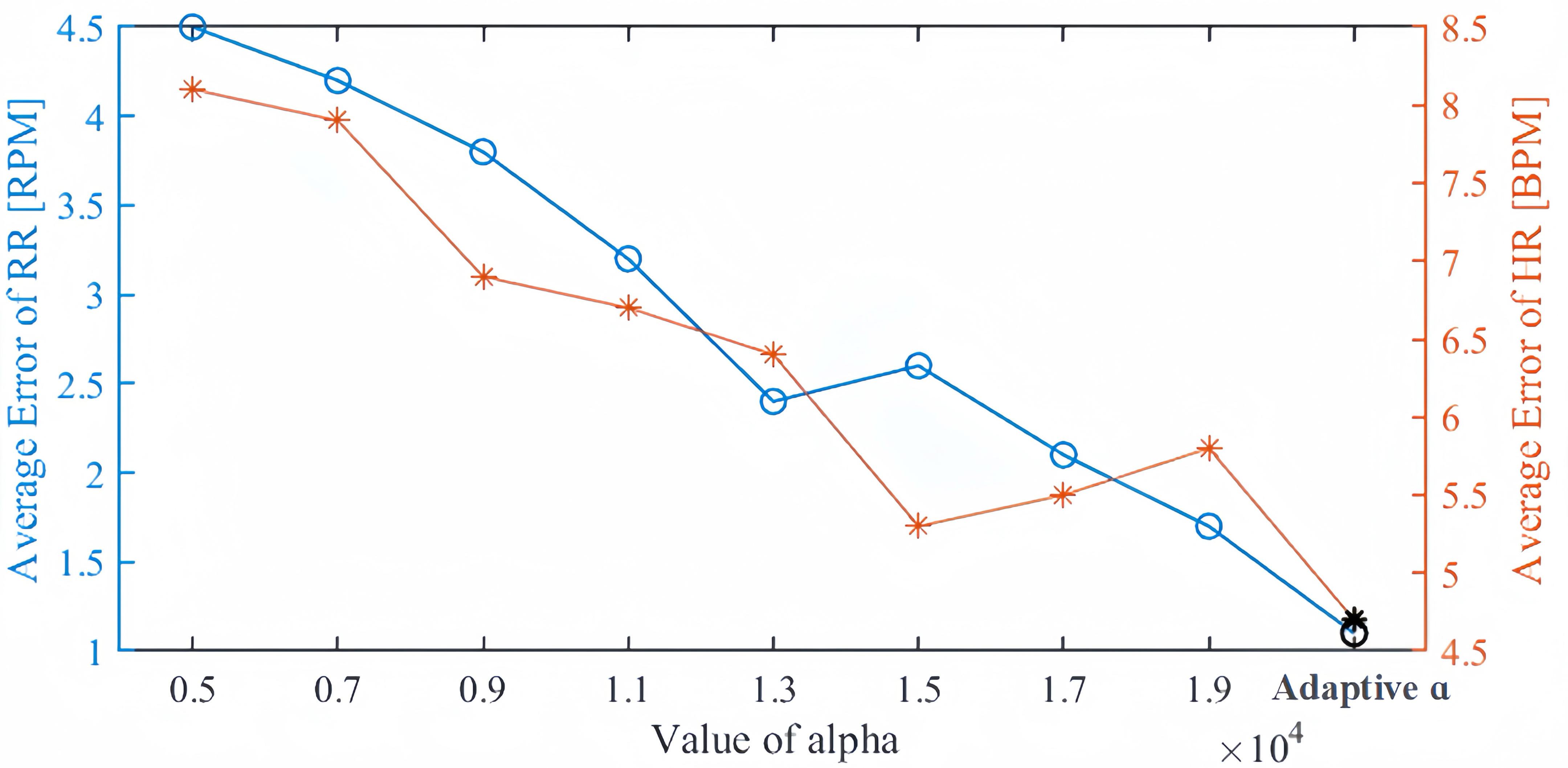}
	\caption{Average errors of vital-sign sensing with different values of $\alpha$.}
	\label{alpha}
\end{figure}

\begin{figure}[h]
	\centering
	\includegraphics[width=\columnwidth]{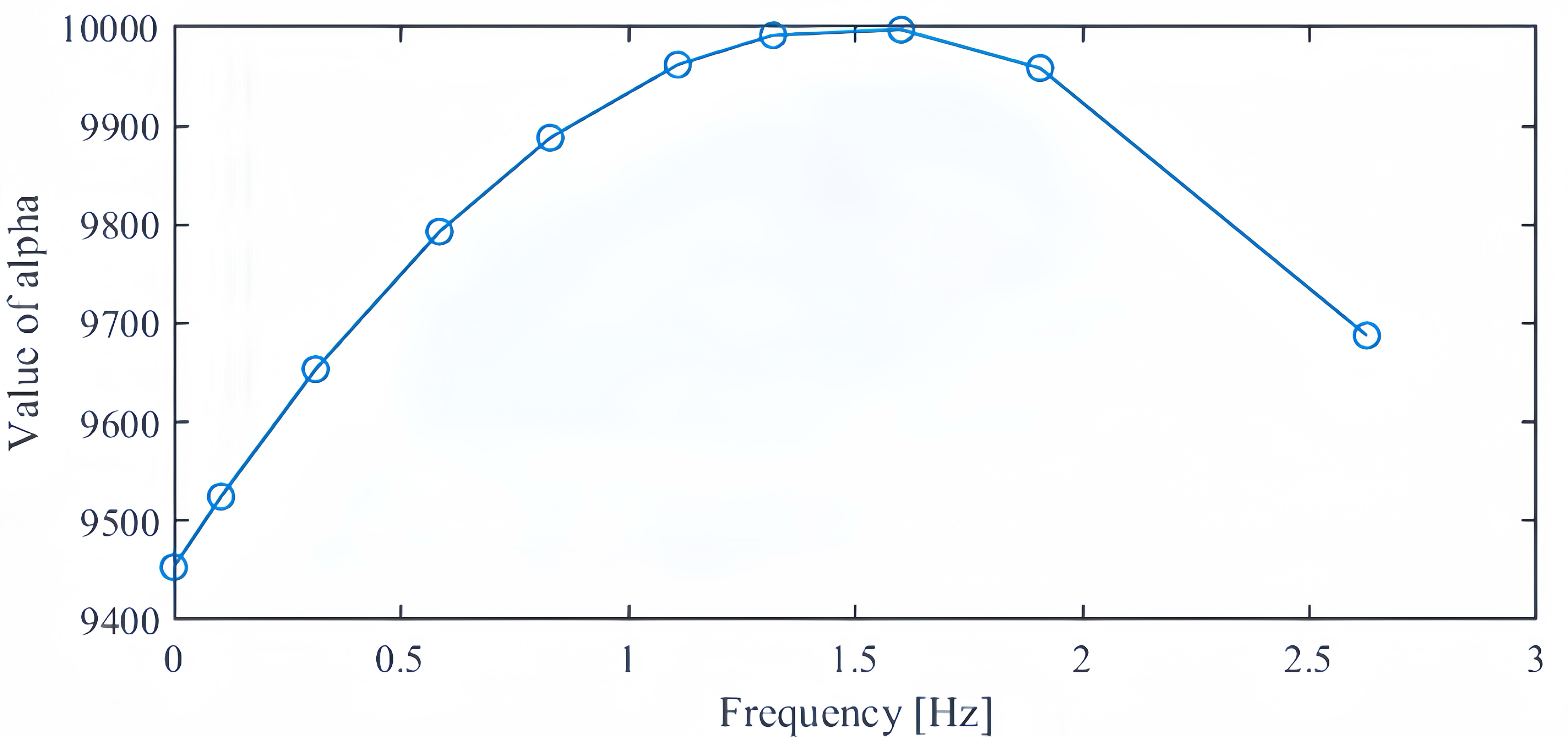}
	\caption{The value of $alpha$ of different IMFs in the improved VMD algorithm.}
	\label{alpha2}
\end{figure}

Fig. \ref{alpha} illustrates the errors of RR and HR estimation with various commonly-used values of the penalty factor $\alpha$ in VMD algorithms. It is evident that utilizing the proposed adaptive $\alpha$ achieves the best performance for vital signs estimation, highlighting the efficiency of the adaptive $\alpha$. Specifically, the average error of estimating RR with an adaptive penalty factor is 0.6 RPM lower than the error when $\alpha$ is set to 1.9e3. Furthermore, the adaptive penalty brings more performance improvement for the HR estimation task, with the average error of HR estimation being 0.7 BPM lower than that when $\alpha$ is 1.5e4. Meanwhile, Fig. \ref{alpha2} illustrates the final value of $\alpha$ for each IMF when there are 10 IMFs to be optimized in the VMD algorithm. It can be observed that the IMF whose center frequency is close to the reference frequency will have a larger $\alpha$. As a consequence, the extracted physiological signals could have a narrower bandwidth, and the noise at other frequencies can be filtered out.

\subsubsection{Filter-based IMF}
\begin{figure}[h]
	\centering
	\includegraphics[width=\columnwidth]{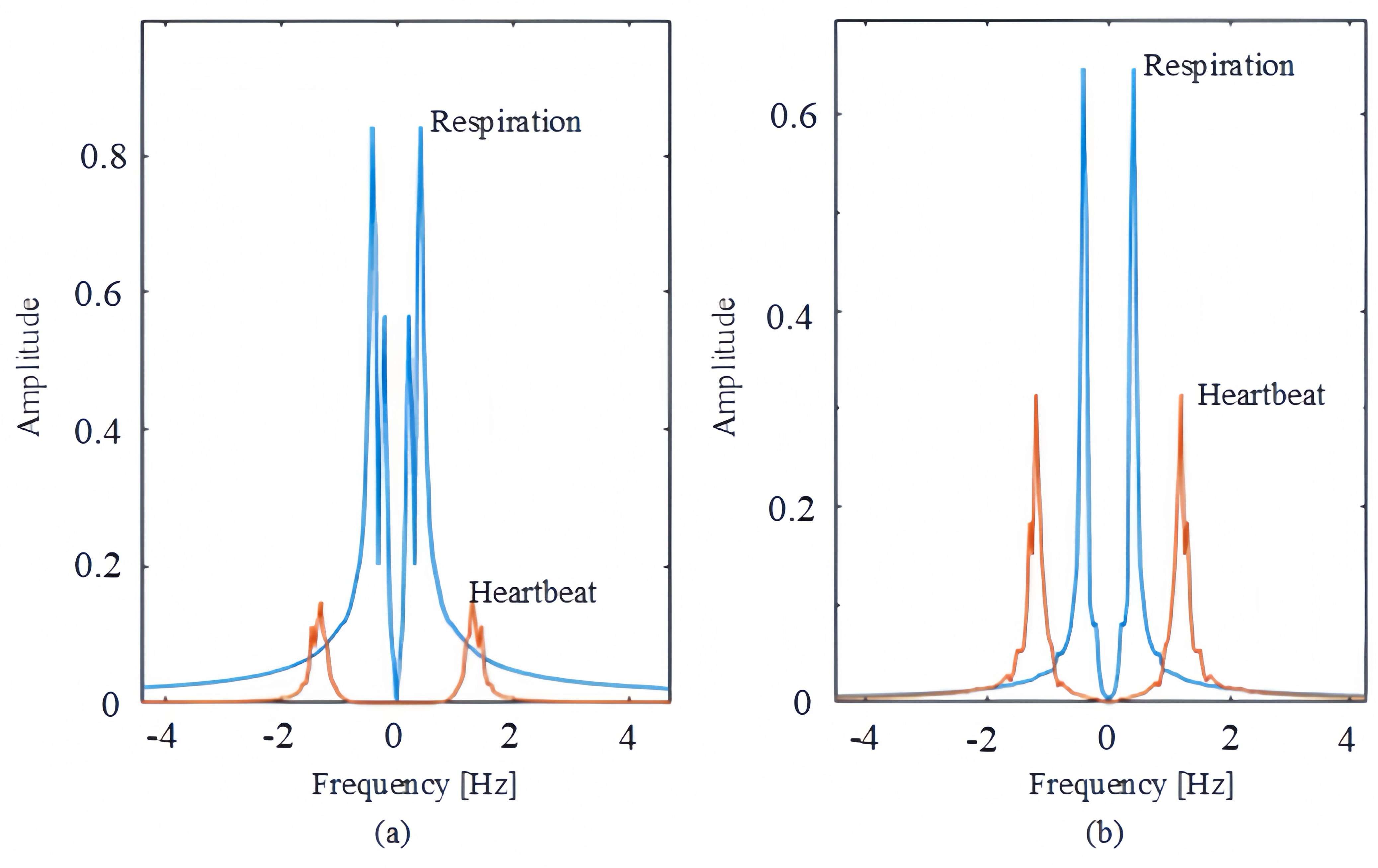}
	\caption{The spectrum of the respiration and heartbeat signal using (a) the conventional VMD and (b) the improved VMD.}
	\label{spectrum}
\end{figure}

Fig. \ref{spectrum} presents the frequency spectrum of the extracted respiration signal and heartbeat signal. From this figure, we can find that implementing the two filters in Eq. (\ref{e4}) on IMFs results in less spectrum overlap between the two physiological signals. This reduction in overlap diminishes the likelihood of drowning the heartbeat signal in the respiration signal. Additionally, Table \ref{t2} presents the RR and HR estimation errors using the filter-based VMD and the conventional VMD. It can be observed that the filter-based strategy reduces the errors of the VMD algorithm for RR and HR estimation.

\subsubsection{The Number of IMFs}

\begin{figure}[h]
	\centering
	\includegraphics[width=\columnwidth]{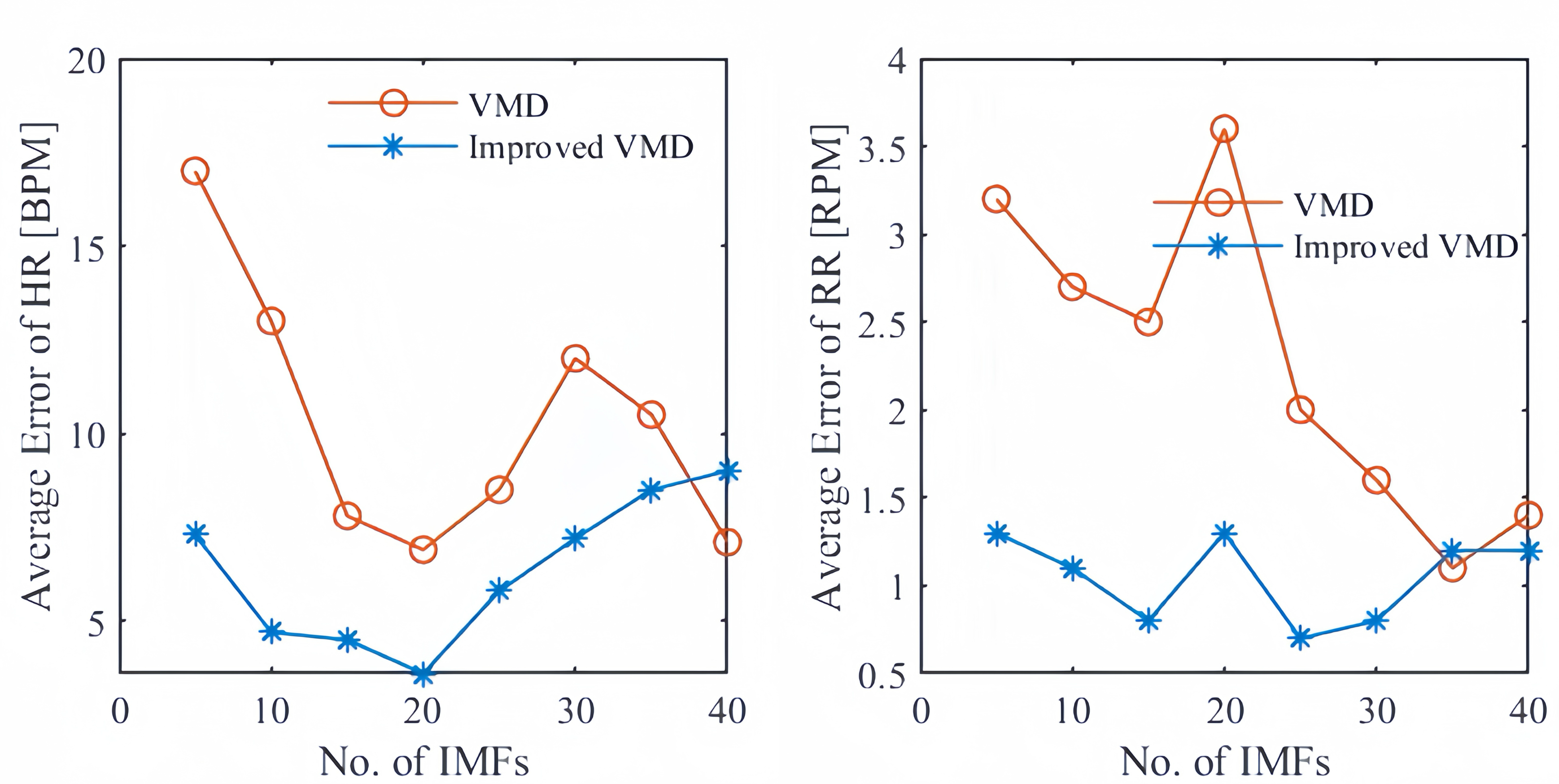}
	\caption{Average errors of vital-sign sensing with different number of IMFs: (a) the average error of HR estimation, and (b) the average error of RR estimation.}
	\label{IMFS}
\end{figure}

The performance variations of the proposed VMD algorithm with different numbers of IMFs are illustrated in Fig. \ref{IMFS}. From this figure, it is observed that the performance of both RR and HR estimation gradually improves and then stabilizes as the number of IMFs increases. However, the complexity of the improved VMD algorithm significantly increases when the number of IMFs is larger, thereby increasing the computational burden and limiting the real-time performance of the algorithm. In comparison with the conventional VMD algorithm, the proposed improved VMD approach is capable of achieving the same performance using fewer IMFs. For instance, as shown in Fig. \ref{IMFS}(a), when there are 10 IMFs, the HR estimation error of the improved VMD is approximately 8 BPM lower than that of the baseline VMD algorithm. Therefore, it can be concluded that the proposed improved VMD algorithm can achieve accurate physiological signal extraction by using fewer IMFs, significantly enhancing the efficiency and real-time capability of the vital-sign sensing system.

\subsection{Impact of Beamforming Empowered by the STC RIS}
\begin{figure}[h]
	\centering
	\includegraphics[width=\columnwidth]{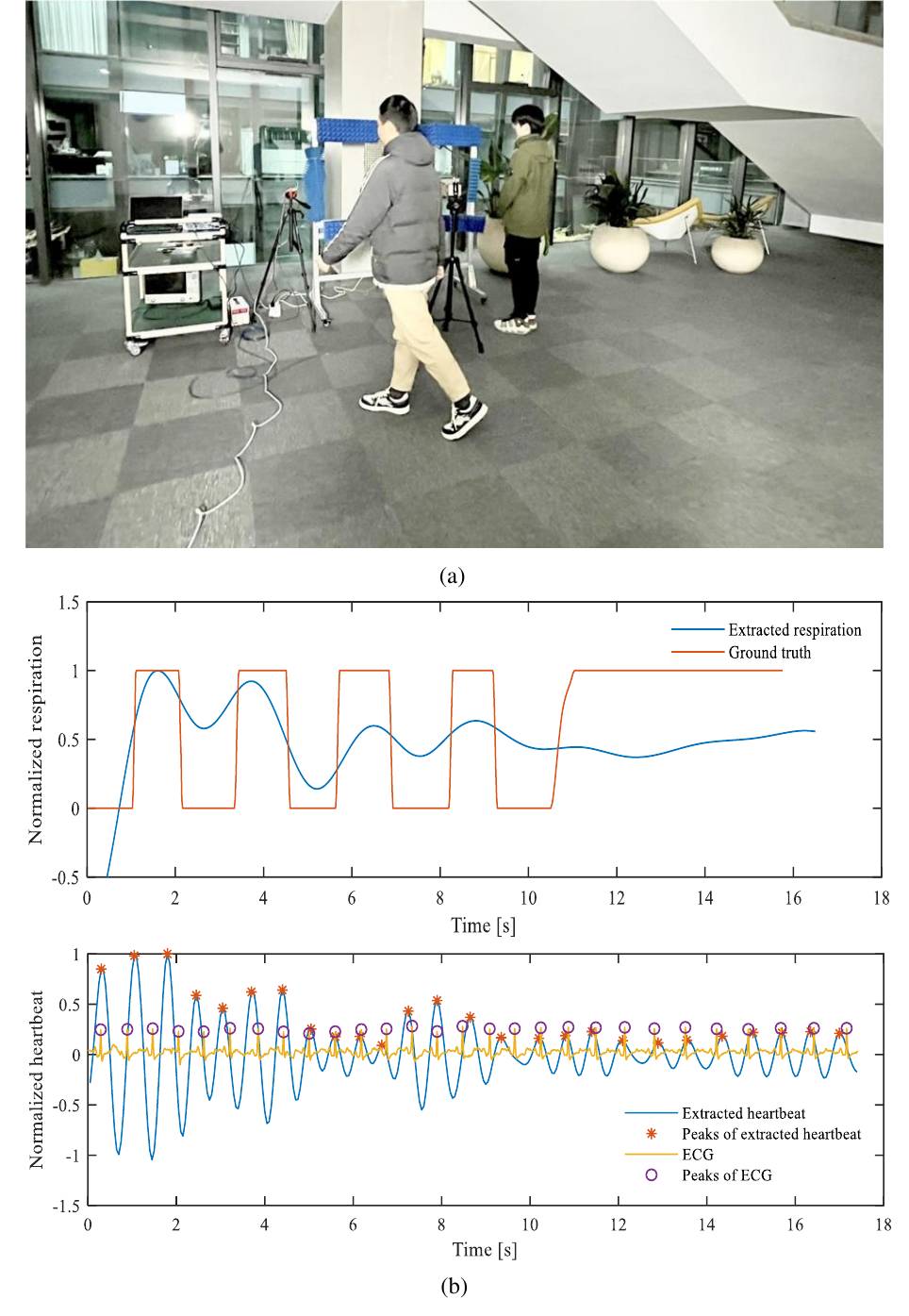}
	\caption{vital-sign sensing results in the scenario where someone is walking by.}
	\label{beamforming}
\end{figure}

In this subsection, we investigate the performance of the vital-sign sensing system in the scenario where there is a passerby walking by the person being monitored, as shown in Fig. \ref{beamforming}(a). When there is another person walking through the AoI, his movements also impact the propagation of sensing signals, making it challenging to extract the target person's vital signs. To address this challenge, we employ the beamforming strategy empowered by the STC RIS to reduce interference at the physical layer. Specifically, by independently controlling each meta-atom unit of the STC RIS, the EM waves reflected by the RIS could be focused on the subject's chest, thereby reducing the influence of irrelevant individuals on sensing signals. As shown in Fig. \ref{beamforming}(b), with the STC RIS-based beamforming, the proposed system can accurately monitor the vital signs of the target individual under the interference of a passerby, with an average error of 0.8 RPM for RR estimation and an error of 5.0 BPM for HR estimation, respectively.  
\subsection{Impact of Distance Between the RIS and Individuals}

\begin{figure}[h]
	\centering
	\includegraphics[width=0.8\columnwidth]{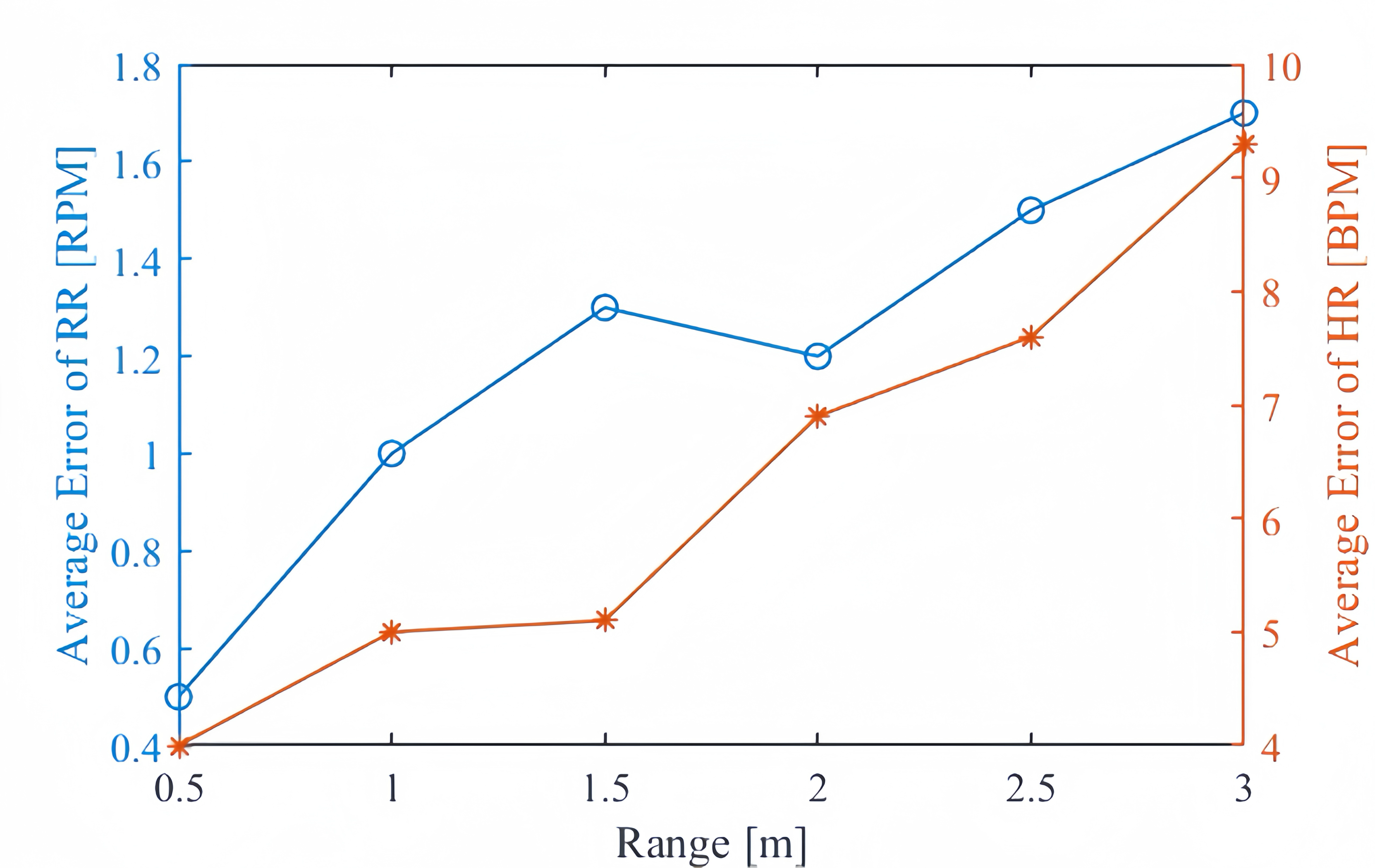}
	\caption{Errors of RR and HR estimation with different distances between the RIS and the human individual.}
	\label{range}
\end{figure}

Furthermore, the impact of the distance between the STC RIS and the individual being monitored is also investigated. As shown in Fig. \ref{range}, with the distance increasing, vital-sign sensing becomes more challenging, resulting in larger errors for RR and HR estimation. Moreover, the performance of HR estimation drops more rapidly than that of RR estimation. The reasons can be explained as follows: On one hand, the impact of human heartbeat on signal propagation is generally weak. Meanwhile, long propagation distance between the RIS and the human target leads to rapid signal attenuation. As a result, the heartbeat signal becomes too weak or even drowned by noise at the receiving terminal, and thus the HR estimation accuracy decreases dramatically with the distance increasing.

\section{Conclusion}
This article proposed a multiperson detection and vital-sign monitoring system empowered by the STC RIS. Taking advantage of the ability to manipulate EM waves in both the spatial and frequency domains, the proposed system was able to detect human targets while simultaneously monitoring their respiration and heartbeat. The RR and HR of each detected individual were jointly estimated using an improved VMD algorithm proposed in this study. To validate the effectiveness of our multiperson passive sensing algorithm, we developed a prototype, and experimental results showed that the proposed system could detect up to 4 persons and accurately monitor their vital signs. Specifically, the errors in RR and HR estimation by using the improved VMD algorithm were below 1 RPM and 5 BPM, respectively. Furthermore, the beamforming controlled by the STC RIS enabled the system to filter out the noises caused by other objects or individuals at the physical layer, thereby improving the SNR of physiological signals and ensuring accurate sensing. In the future, we aim to further investigate multiperson vital-sign sensing with the aid of RISs, particularly in scenarios involving individuals in motion, such as walking or running. This will further broaden its application potential in various real-world contexts.

\bibliographystyle{IEEEtran}
\bibliography{refs}

\newpage

\vfill

\end{document}